\documentclass[sigconf, nonacm]{acmart}

\usepackage{subcaption}
\usepackage{endnotes,microtype,xspace,graphicx,fancyvrb,multirow} 
\usepackage[linesnumbered, ruled, vlined,resetcount]{algorithm2e} 
\usepackage{algpseudocode}
\usepackage{soul}
\usepackage{xcolor}           %
\usepackage[normalem]{ulem}
\usepackage{setspace}

\newtheorem{definition}{Definition}

\usepackage{balance}
\usepackage{newfloat}
\DeclareFloatingEnvironment[fileext=frm,placement={!ht},name=Frame]{myfloat}
\makeatletter
\newcommand{\removelatexerror}{\let\@latex@error\@gobble}
\makeatother

\usepackage{latexsym}
\usepackage{amsfonts}
\usepackage{amsmath}
\usepackage{amssymb}
\usepackage{color}
\usepackage{epsfig}
\usepackage{xspace}
\usepackage{graphicx}
\usepackage{balance}
\usepackage{hhline}
\usepackage{float}
\usepackage{xcolor}
\usepackage{tikz}

\usepackage{epsfig}
\usepackage{multirow}
\usepackage{url}

\usepackage{enumitem}
\setlist{topsep=0pt,noitemsep} \setitemize[1]{label=$\circ$}

\newcommand{\ei}{\end{itemize}\vspace{1ex}}

\newcommand{\ee}{\end{enumerate}\vspace{1ex}}
\newcommand{\beqn}{\begin{eqnarray}}
\newcommand{\eeqn}{\end{eqnarray}}

\newcommand{\stitle}[1]{\vspace{1.2ex}\noindent{\bf #1}}
\newcommand{\etitle}[1]{\vspace{1.2ex}\noindent{\em\uline{#1}}}

\newcommand{\ie}{i.e.,\xspace}

\renewcommand{\smallskip}{\vspace{0.6ex}}

\newcommand{\kw}[1]{{\ensuremath {\mathsf{#1}}}\xspace}

\newcounter{ccc}

\newcommand{\eop}{\hspace*{\fill}\mbox{$\Box$}}     %
\newcounter{example}%
\renewcommand{\theexample}{\arabic{example}}
\newenvironment{example}{
        \vspace{1.2ex}
        \refstepcounter{example}
        {\noindent\bf Example \theexample:}}{
        \eop\vspace{1.2ex}}

\newlist{myitemize}{itemize}{3}
\setlist[myitemize,1]{label=$\circ$,leftmargin=3.8ex}
\setlist[myitemize,2]{label=$\bullet$,leftmargin=3.8ex}
\setlist[myitemize,3]{label=$\diamond$, leftmargin=3.2ex}

\newcommand{\mei}{\end{myitemize}\vspace{0.6ex}}

\usepackage{tikz}

\newcommand{\oursys}{\kw{ACTIVE}\xspace}
\newcommand{\ddeg}{{DEG}\xspace}

\newcommand{\pfrepair}{{CA-PFR}\xspace}
\newcommand{\acprune}{{ACAR}\xspace}

\newcommand{\circlednum}[1]{%
  \tikz[baseline=(char.base)]{
    \node[circle, draw, line width=0.4pt,
          inner sep=1.2pt, font=\footnotesize\bfseries] (char) {#1};
  }%
}

\usepackage{xcolor}
\usepackage[normalem]{ulem}

\newcommand{\delopsonerbrmin}{22.64}

\newcommand{\delopsonenrbrmin}{1.64}
\newcommand{\delopsonenrbrmax}{2.06}
\newcommand{\delcmponerbrmin}{8.72}

\newcommand{\delcmponenrbrmax}{2.53}

\newcommand{\inscmponemax}{1.43}

\newcommand{\insopsonemax}{1.43}

\newcommand{\delrecallonenrbrmin}{0.21\%}

\usepackage{comment}
\AtBeginDocument{\color{black}\hypersetup{allcolors=black}}

\begin{document}

\pagestyle{plain}
 \clearpage
\setcounter{page}{1}

\title{
ACTIVE: A Dynamic Pareto-Frontier-Based Unified Index for Dual-Vector Query
}

\settopmatter{authorsperrow=4}

\author{Shengyuan Lin}
\affiliation{
  \institution{%
  Northeastern Univ., China}
}
\email{linsy5847@gmail.com}

\author{Shufeng Gong}
\affiliation{
  \institution{%
  Northeastern Univ., China}
}
\email{gongsf@mail.neu.edu.cn}

\author{Zheng Wang}
\affiliation{
  \institution{%
  Northeastern Univ., China}
}
\email{wangzheng29a@163.com}

\author{Feng Yao}
\affiliation{
  \institution{%
  Northeastern Univ., China}
}
\email{yaofeng@stumail.neu.edu.cn}

\author{Shifu Li}
\affiliation{
  \institution{Huawei Tec. Co., Ltd}
}
\email{lishifu@huawei.com}

\author{Yanfeng Zhang}
\affiliation{
  \institution{%
  Northeastern Univ., China}
}
\email{zhangyf@mail.neu.edu.cn}

\author{Ge Yu}
\affiliation{
  \institution{%
  Northeastern Univ., China}
}
\email{yuge@mail.neu.edu.cn}

\begin{abstract}
Existing approximate nearest neighbor search (ANNS) methods generally assume that each entity has a single vector. In many real-world applications, however, an entity is associated with more than one vector, %
and queries may jointly consider two attributes. %
With different queries assigning different weights to each attribute, %
Pareto-frontier-based indices provide an effective solution for dual-vector retrieval by connecting to approximate nearest neighbors under all possible query preferences. However, existing Pareto-frontier-based approaches suffer from substantial inefficiencies in both index construction and dynamic updates, limiting their practical applicability.

To address these issues, we propose \oursys, a dynamic Pareto-frontier-based DVQ index that supports efficient construction and updates while maintaining high retrieval performance under varying query preferences. It employs angle-constrained active-range computation and connectivity-loss-driven adaptive repair strategies to accelerate node insertion and deletion, respectively.
{Extensive experiments show that, at comparable (or slightly below) search accuracy, \oursys improves insertion throughput by $1.20$--$1.43\times$ over DEG, and deletion throughput by $1.64$--$2.06\times$ over neighbor-reconnection-based repair and by $22.64$--$39.51\times$ over reinsertion-based repair. %
}%

\end{abstract}

\maketitle

\section{Introduction}
\label{sec:intro}

\begin{figure}[tbp]
    \centering
    \includegraphics[width=0.99\linewidth]{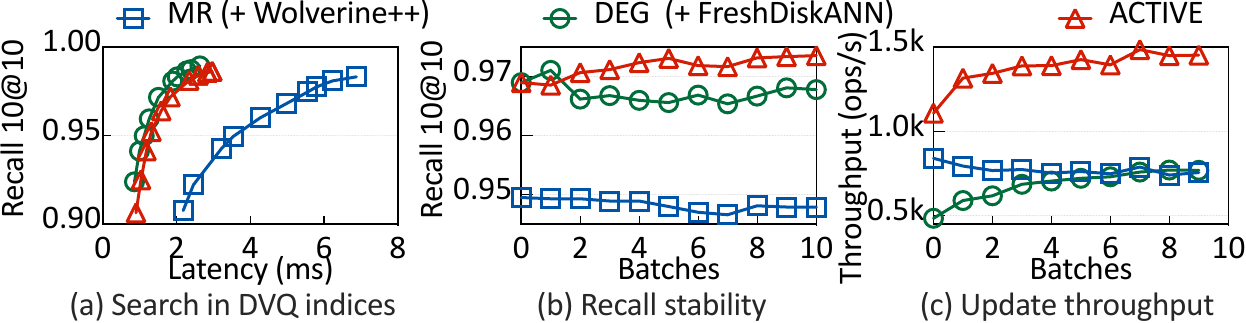}%
    \vspace{-0.15in}
    \caption{{Performance of DVQ indices under static search and dynamic updates on the OpenImage dataset, where ``MR'' denotes multi-index retrieval.}}
    \vspace{-0.25in}
    \label{fig:motivation}
\end{figure}

{

}

Approximate nearest neighbor search (ANNS) over high-dimensional vectors has been widely adopted in applications such as recommendation systems~\cite{RecSys1,RecSys2,RecSys3,RecSys4} and information retrieval~\cite{InfoRet1,InfoRet2,InfoRet3,InfoRet4}.
In these applications, the unstructured attributes of an entity, such as image, text, or audio, are embedded into vectors, and objects whose vectors are close (similar) to the query vector are retrieved. Traditional ANNS primarily focuses on single-vector retrieval~\cite{HNSW,NSG,DiskANN,Starling,IVFADC,LSH,SPANN,spfresh}, \ie entities are searched based on only \textit{one} unstructured (vectorized) attribute. %

However, many real-world retrieval tasks require jointly considering more than one vectorized attribute~\cite{dvqs1,dvqs2,dvqs3,dvqs4,dvqs5,dvqs6,milvus}, with query-specific preferences assigned to each attribute. %
{In this paper, we focus on dual-vector retrieval with exactly two attribute vectors, which is common in applications with complementary modalities, such as image--text {of products} on e-commerce platforms~\cite{milvus,deg,huawei2025multivector}, spatial--textual data {of posts} on social platforms~\cite{dvqs4}, and video--text {of videos} on short-video platforms~\cite{deg,dvqs5}.}
{Given a query with two vectors, the distance between the query and an object is the weighted sum of their distances on the two vectors, where the query-specific preference determines the weight of each attribute distance.
As query preferences vary, the distance between the same pair of entities changes accordingly, making efficient and accurate ANN retrieval particularly challenging.} %
Furthermore, real-world datasets are inherently dynamic, with entities continuously being inserted and deleted.
Such dynamic updates pose significant challenges to efficient index-based retrieval methods.

\begin{example}
{%
In product search on e-commerce platforms, consider two users who upload images of a smartphone and provide textual descriptions, %
such as ``a smartphone with long battery life and excellent photography''.
The first user aims to retrieve a smartphone that visually resembles the uploaded image, whereas the second user prioritizes the specific performance attribute.
Accordingly, the first user assigns a larger weight to the image-vector distance, while the second user assigns a larger weight to the textual-vector distance when computing similarity based on the image and textual embeddings.}  
Even when the two users provide the same phone image and textual description, different preferences can lead to different retrieval results.
Meanwhile, vendors continuously update product listings due to inventory changes, ensuring accurate retrieval results for users. This makes the dynamic updating of dual-vector data indispensable.
Thus, the DVQ problem requires efficient and accurate retrieval under arbitrary query preferences while also supporting efficient dynamic updates.
\end{example}

{
\stitle{Motivation}. 
Existing DVQ approaches~\cite{milvus,deg,multi-recall,fixed-alpha} %
mainly build graph-based ANNS indices over dual-vector data to enable efficient retrieval. However, they cannot simultaneously achieve high retrieval accuracy and high update efficiency.
In general, these approaches can be divided into two categories.

\etitle{1) Preference-specific multi-indices.}
These approaches~\cite{milvus,multi-recall,fixed-alpha} build single-vector indices by fixing the query preference in advance. 
{A representative solution is multi-index retrieval (MR)~\cite{milvus,multi-recall}, which builds a separate index {for each attribute vector}.
It searches the two indices independently, merges the retrieved candidates, and subsequently reranks them by computing the exact weighted distance.}
Since they assume that object distances are pre-determined, they suffer from limited retrieval quality under arbitrary query preferences, as shown in Figure~\ref{fig:motivation}a. 

\etitle{2) Preference-agnostic unified indices.}
Unlike existing single-vector indices~\cite{DiskANN,HNSW,NSG} that select similar vectors of each node as their neighbors,
this approach~\cite{deg} connects 
approximate nearest neighbors across all possible query preferences,
ensuring that varying query preferences can be accommodated during search.
To achieve this, approximate Pareto frontiers are utilized %
as the neighbors since they remain competitive under diverse preference settings, thereby enabling the selected neighbors to cover varying query preferences.
As a result, Pareto-frontier-based indices provide a more suitable foundation for query-dependent similarity search in DVQ.
However, existing unified indices do not support efficient dynamic updates because repairing Pareto frontiers is more complex than repairing neighbors in single-vector indices~\cite{FreshDiskANN,wolverine,greator}.

}

\stitle{Our goal}. 
In this work, we aim to design a dynamic Pareto-frontier-based DVQ index that supports efficient construction and updates while maintaining high retrieval performance. %

\stitle{Challenges}.
Designing a preference-agnostic unified index is challenging for two main reasons.

\etitle{1) Expensive Edge Active-Range Computation in Construction.} %
{
Pareto-frontier-based indices~\cite{deg} select neighbors to cover different query preferences, but not every selected edge is useful under every preference.
Thus, each edge needs to precompute its active ranges of the preference, allowing it to be traversed only when needed and avoiding redundant distance computations.
Employing Relative Neighborhood Graph (RNG) pruning~\cite{deg} enables the computation of such active ranges, however, its expensive time complexity introduces substantial additional computational overhead. }
Therefore, designing an efficient active-range computation method remains challenging. 

\etitle{2) Inefficient Approximate Pareto Frontier Repair in Update.} %
{
In Pareto-frontier-based indices, repairing the Pareto-frontier neighbors of affected nodes (i.e., in-neighbors of deleted nodes) after deletions is critical to preserving graph connectivity.
Reinsertion-based global repair strategies~\cite{IPGM,pgvector,HNSW} rediscover the neighbors of affected nodes by traversing from entry nodes, causing many distance computations and high update overhead.
However, directly applying single-vector local repair strategies~\cite{FreshDiskANN,greator} to Pareto-frontier-based DVQ indices is ineffective, as validated in Figure~\ref{fig:motivation}b and \ref{fig:motivation}c, since they cannot guarantee that Pareto frontiers can be identified within the local neighborhood of DVQ indices.}
Therefore, designing an efficient repair strategy for approximate Pareto frontiers while maintaining high retrieval accuracy remains challenging.

\stitle{ACTIVE}. 
{The above challenges cannot be
directly addressed by existing techniques. Starting from these application-level challenges,} we propose \oursys, a dyn\underline{A}mi\underline{C} Pareto-fron\underline{T}ier-based un\underline{I}fied D\underline{V}Q ind\underline{E}x that supports dynamic updates while maintaining high retrieval quality under arbitrary query preferences.
\oursys is built upon two techniques.

To accelerate index construction, we {introduce the Pareto angle} and develop an \underline{a}ngle-\underline{c}onstrained \underline{a}ctive \underline{r}ange computation algorithm, \acprune, {based on the locality distribution of Pareto angles.}
In \acprune, each neighbor is compared only with other neighbors that satisfy the angle constraint under the triangle inequality. 
By restricting unnecessary comparisons, \acprune reduces construction overhead while preserving effective navigation quality of the DVQ index across diverse query preferences.%

For efficient updates, we
{observe a high overlap between the original neighbors and the %
repaired neighbors of affected nodes}
and introduce \pfrepair, a connectivity-loss-driven adaptive repair strategy that locally repairs approximate Pareto frontiers.
We provide three methods, namely historical-Pareto-frontier, affected-layer, and multi-layer repair, according to the severity of connectivity degradation of affected nodes.
This allows us to restore the connectivity of the graph without full re-insertion, significantly reducing distance computations and improving update throughput.

In summary, the main contributions of this work are:
\begin{itemize}[leftmargin=0.1in]

\item[$\bullet$] An {observation of Pareto-angle locality motivates an angle-constrained active range computation algorithm,} which accelerates active range computation %
($\S$~\ref{sec:construct}).

\item[$\bullet$] An 
{observation of high overlap between original and %
repaired neighbors of affected nodes
motivates an incremental repair method.
We propose a connectivity-loss-driven adaptive repair strategy,} which locally repairs approximate Pareto frontiers according to the severity of connectivity degradation ($\S$~\ref{sec:deletion}).

\item[$\bullet$] {Extensive experiments on diverse datasets show that \oursys achieves higher update throughput than state-of-the-art baselines while maintaining a competitive trade-off between query accuracy and efficiency ($\S$~\ref{sec:expr}).}

\end{itemize}

{%
}

{
}%

{

}

{
}

{%
}

{
}
{%

}

{%

}

{%

}

\vspace{-0.1in}
\section{Background And Motivation}
\label{sec:bg}

In this section, we first define the Dual-Vector Query (DVQ) problem and review Pareto-frontier-based DVQ index construction as well as update methods for single-vector graph-based indices. We then analyze their drawbacks and present the motivation of this work.

\vspace{-0.1in}
\subsection{Dual-Vector Query}
\label{sec:bg:graph_DVQ}

Let ${O} = \{o_0, o_1, \dots, o_{N-1}\}$ denote a dataset of $N$ objects, where each object $o_i \in {O}$ is represented by two feature vectors: 
(1) $o_i.x \in \mathbb{R}^{d_x}$, a $d_x$-dimensional vector capturing one attribute, and 
(2) $o_i.y \in \mathbb{R}^{d_y}$, a $d_y$-dimensional vector capturing another attribute.

\stitle{Dual-Vector Query (DVQ)}.
Given a query 
$q = \langle x, y, \alpha, k \rangle$,
where $q.x \in \mathbb{R}^{d_x}$ and $q.y \in \mathbb{R}^{d_y}$ are the query vectors, {$q.\alpha \in [0,1]$ is a weighting parameter provided by upstream applications to quantify user preferences}, and $q.k$ is the number of objects to retrieve.
Ideally, DVQ aims to retrieve the exact top-$k$ nearest objects $R_{{exact}} \subset {O}$ for a query $q$, such that $\forall e \in R_{{exact}}, \forall o \in {O} \setminus R_{{exact}}$, we have $\mathrm{Dist}(q,e) \leq \mathrm{Dist}(q,o)$, where the distance~\cite{deg} between $q$ and an object $o$ is defined as:
\begin{equation}
\label{eq:DVQ_dist}
\mathrm{Dist}(q, o) = q.\alpha \cdot \delta_x(q, o) + (1-q.\alpha) \cdot \delta_y(q, o),
\end{equation}
where
$\delta_x(q, o) = \frac{\|q.x - o.x\|_2}{x_{\max}}$ and
$\delta_y(q, o) = \frac{\|q.y - o.y\|_2}{y_{\max}}$
are the normalized distances for each vector.
{A larger $\alpha$ means the user prefers the $x$-attribute, while a smaller $\alpha$ means the user prefers the $y$-attribute.
In real applications, $\alpha$ can be learned and tuned based on the user's query history and the entities ultimately selected.}

However, the retrieval of $R_{{exact}}$ requires exhaustive search over the entire dataset,
which becomes computationally expensive and incurs high search latency~\cite{ANNS}.
To address this, DVQ in practice retrieves $k$ approximate nearest results
$R_{DVQ}$, achieving a tradeoff between retrieval accuracy and search efficiency~\cite{ANNS,deg}.
To quantify the quality of the approximate results $R_{DVQ}$, recall~\cite{recall_1,recall_2} is commonly used, defined as
$Recall@k = \frac{| R_{DVQ} \cap R_{exact} |}{k}$,
where $R_{exact}$ is the ground-truth set of the $k$ closest objects to $q$. The goal of DVQ is to maximize recall while retrieving results efficiently.

\begin{figure}[tbp]
    \vspace{-0.5in}
    \centering
    \includegraphics[width=.99\linewidth]{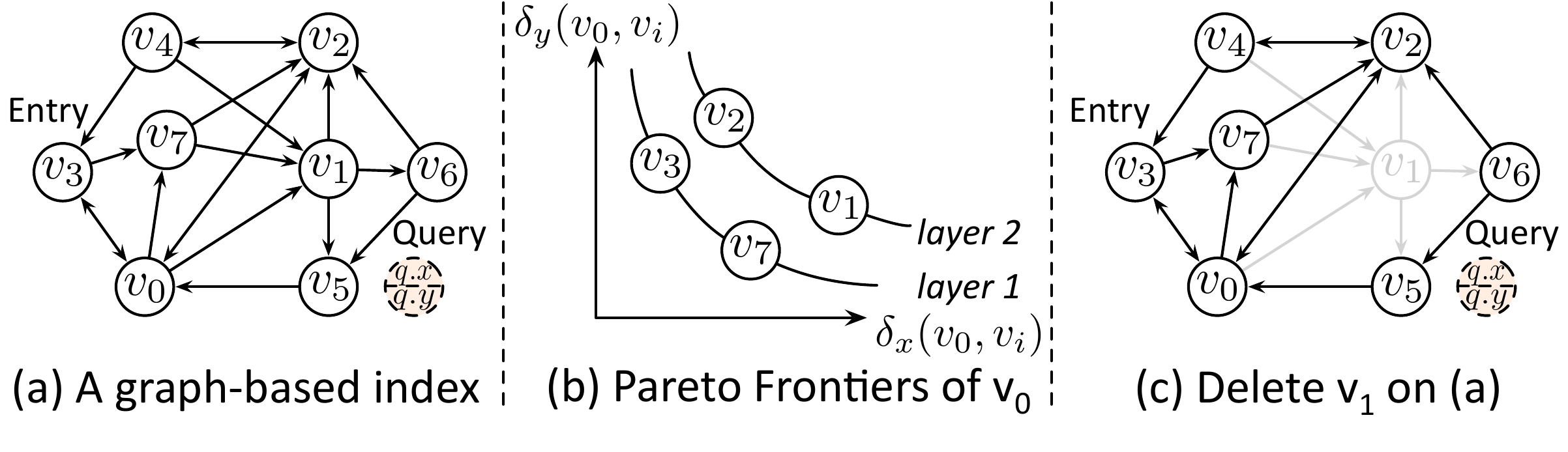}
    \vspace{-0.2in}
    \caption{An example of a graph-based DVQ index. Each node maintains at most $R=4$ neighbors.
    The dashed circle indicates a dual-vector query, $v_3$ is the entry node. %
    }
    \vspace{-0.15in}
    \label{fig:bg_index}
\end{figure}

\vspace{-0.1in}
\subsection{Pareto-Frontier-Based Index Construction}
\label{sec:bg_build}

{%
Due to diverse query $\alpha$ in DVQ, Pareto-frontier-based indices select the Pareto frontiers of each node as its out-neighbors, which can nearly cover the approximate nearest neighbors for all possible $\alpha$. %
}

\stitle{Pareto Frontier}~\cite{pareto-frontier}. 
The construction of Pareto-frontier-based indices can be viewed as finding the Pareto frontier for each node.
Given a node $v \in V$ for which we aim to find its neighbors, each candidate node $p \in V \setminus \{v\}$
induces a distance pair relative to $v$, $\big(\delta_x(v,p), \delta_y(v,p)\big)$.
For any $p, q \in {V} \setminus \{v\}$, we say that $p$ \textit{dominates} $q$, denoted by $p \prec q$, if and only if
(1) $\forall i \in \{x,y\}, \ \delta_i(v,p) \le \delta_i(v,q)$, and
(2) $\exists j \in \{x,y\}, \ \delta_j(v,p) < \delta_j(v,q)$.
The Pareto frontier of $v$ is defined as $PF_v \subseteq {V} \setminus \{v\}$, where $\forall p \in PF_v$, $\nexists q \in {V}\setminus\{v\}, q\prec p$, \ie the neighbors of $v$ in $PF_v$ cannot be dominated by any node.

\stitle{Multi-layer Pareto Frontiers}.
In Pareto-frontier-based indices, only using the Pareto frontier of each node as its out-neighbors leads to an extremely sparse graph index.
This is because, in a dataset with millions of objects, the size of the Pareto frontier is typically small (e.g., the average size is ~8.26 on CC3M)
, while hundreds of candidate nodes are usually required to select as the final out-neighbors.
To address this, the multi-layer Pareto frontiers $\{PF^1_v, \dots,PF^l_v\}$ are selected
as the candidate out-neighbors of each node $v$, where $l$ is the number of layers bounded by the candidate pool size $ef_{construct}$, as shown in Figure~\ref{fig:bg_index}b.
{Neighbors in the first layer correspond to the global Pareto frontier. Each subsequent layer is recursively constructed by removing the {current Pareto frontier}
and extracting a new Pareto frontier from the remaining candidates, referred to as \textsc{FindPF}~\cite{skyline}.}

\stitle{Approximate Pareto Frontier Search}.
However, the construction of exact multi-layer Pareto frontiers is computationally expensive, as it requires computing distances to all other nodes in the dataset for each node.
Therefore, approximate Pareto frontiers are typically used as candidate neighbors for each node, which are constructed using the Greedy Pareto Frontier Search (GPS) to expand candidate neighbors
layer by layer
(see Algorithm~\ref{alg:gps}).
{Specifically, GPS iteratively expands unexplored nodes in the lowest layer of Pareto frontiers containing them until no unexplored nodes remain.
Finally, the candidate set ${\mathcal{C}}$ is selected as the approximate Pareto frontier.}

\begin{myfloat}[t]
\vspace{-0.5in}
\removelatexerror
\setstretch{0.8}
\begin{algorithm}[H]
\caption{GPS($G, q, EP, ef_{{construct}}$)}
\label{alg:gps}
\KwIn{Graph $G=(V,E)$, query node $q$, entry node set $EP$, candidate pool size $ef_{{construct}}$}
\KwOut{Multi-layer Pareto frontiers $\mathcal{C}=\{PF^1_q, \dots, PF^l_q\}$ of $q$}

$\mathcal{C} \gets \textsc{FindPF}(EP,ef_{construct})$\;

\While{$|\mathcal{C}| < ef_{{construct}}$}{
    $NNS \gets$ unexplored nodes from the lowest layer in $\mathcal{C}$ that contains such nodes\;
    \If{$NNS = \emptyset$}{
        {break}\;
    }
    Add unvisited out-neighbors of nodes in $NNS$ to $\mathcal{C}$\;
    
    $\mathcal{C} \gets \textsc{FindPF}(\mathcal{C}, ef_{{construct}})$\;
}
\Return{$\mathcal{C}$}\;
\end{algorithm}
\vspace{-0.25in}
\end{myfloat}

\stitle{Active Range (AR) Computation}.
{
Since the out-neighbors of each node cover the nearest neighbors under varying query preferences, a significant portion of the outgoing edges need not be traversed for a given $\alpha$ value.
To accelerate query processing, we assign preference intervals
(\ie \emph{active ranges})
to each edge, enabling the system to selectively traverse only the edges whose active ranges include the query $\alpha$.
}
However, computing the active ranges of each edge is difficult \cite{deg}. DEG adopts an RNG-inspired pruning strategy similar to those used in single-vector graph indices. %
Specifically, for each candidate neighbor in $\mathcal{C}$ corresponding to an outgoing edge, it evaluates the distance to every other candidate neighbor and computes an interval of the parameter $\alpha$ over which the edge becomes the longest edge in the corresponding triangle configuration. The active range is then obtained by excluding all such intervals from the complete parameter domain $[0,1]$.
This method requires comparing each neighbor of a node with all retained neighbors to determine its active ranges, resulting in a time complexity of $O(R \cdot |\mathcal{C}|)$, where $R$ is the maximum out-degree.

\stitle{Summary}.
Existing Pareto-frontier-based DVQ index construction methods still face performance challenges, as the active range computation introduces substantial computational overhead.

\vspace{-0.1in}
\subsection{Pareto-Frontier-Based Index Updates}
\label{sec:bg_update}

In real-world scenarios, timely updates of indices are essential for search accuracy.
Although no specialized DVQ index has yet been proposed to efficiently support dynamic updates, especially for node deletions, existing update techniques for single-vector graph-based indices are already well-established~\cite{IPGM,pgvector,FreshDiskANN,greator}.
Next, we review representative update methods in existing single-vector graph-based indices. The insertion procedure can follow the same logic as the construction process. Therefore, we focus our discussion on the deletion case.

Existing repair strategies can be broadly classified into the following two categories:
1) \textit{Reinsertion-Based Repair (RBR)} methods~\cite{HNSW,IPGM,pgvector} treat affected nodes as newly inserted nodes and reinsert them to reconstruct their out-neighbors.
Although these methods can maintain high index quality, the global search incurs significant computational overhead.
2) \textit{Neighbor-Reconnection-Based Repair (NRBR)} methods~\cite{FreshDiskANN,greator} restore graph connectivity through local reconnections, based on the intuition that neighbors of neighbors are likely to remain close to the affected node.
Compared with RBR, NRBR may degrade index quality while achieving higher update throughput.
However, in Pareto-frontier-based DVQ indices, these NRBR methods cannot guarantee that the neighbors of a node's neighbors belong to its Pareto frontiers, making it difficult to maintain high retrieval accuracy.

\begin{figure}[tbp]
     \vspace{-0.5in}
    \centering
\includegraphics[width=0.85\linewidth]{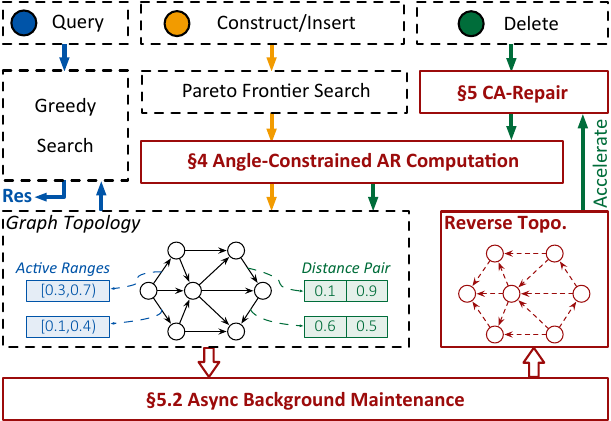}
    \vspace{-0.1in}
    \caption{Workflow of \oursys.}
    \vspace{-0.27in}
    \label{fig:workflow}
\end{figure}

\stitle{Summary}.
{Update strategies for single-vector graph indices provide useful insights for DVQ index updates.
For instance, the insertion operation can largely reuse techniques from incremental construction.
Nevertheless, efficient updates for DVQ indices remain challenging. In particular,
designing an efficient deletion repair strategy that preserves high index quality remains challenging.} 

\stitle{Limitations}. %
Based on the analysis in Section~\ref{sec:bg_build} and \ref{sec:bg_update}, existing Pareto-frontier-based DVQ indices suffer from inefficient index construction and substantial update overhead.

\vspace{-0.05in}
\section{Overview}
\label{sec:overview}

To overcome the above limitations, we design \oursys, as shown in Figure~\ref{fig:workflow}. %
To accelerate index construction,
1) we introduce an \textit{angle-constrained active range computation} ($\S$\ref{sec:construct}) strategy that avoids comparing each candidate neighbor against all retained neighbors by leveraging angular constraints. %
To accelerate index updates,
2) we propose a \textit{connectivity-loss-driven adaptive repair} ($\S$\ref{sec:deletion}) strategy that locally repairs the approximate Pareto frontiers of affected nodes according to the severity of their connectivity degradation. %
Furthermore, %
we incorporate \textit{asynchronous reverse graph maintenance} to efficiently locate %
the affected nodes ($\S$\ref{sec:ca-repair}).

\etitle{Index Construction}.
\oursys employs an incremental construction strategy for each object in the dataset. 
Specifically, when constructing the out-neighbors of node $v$, the candidate set $\mathcal{C}$ is obtained by performing greedy Pareto frontier search over the current index.
Subsequently, \oursys computes the active ranges of nodes in $\mathcal{C}$ using the proposed \acprune and selects the top-$R$ unpruned neighbors in lower layers of Pareto-frontiers as the final out-neighbors $N_{\text{out}}(v)$.
Finally, for each $v' \in N_{\text{out}}(v)$, a reverse edge $(v', v)$ is added to guarantee reachability of $v$ during subsequent searches.

\etitle{Index Update}.
For insertion, \oursys adopts the incremental construction approach to add new nodes into the graph index.
For deletion, \oursys first leverages reverse graph topology to quickly locate the affected nodes. 
Then, \oursys adaptively repairs the affected nodes with three methods, namely historical-Pareto-frontier repair, affected-layer repair, and multi-layer repair, according to the severity of connectivity degradation for each affected node.  
Finally, \oursys computes the active ranges of the candidate neighbors for each affected node and selects the resulting final neighbors.

\etitle{Dual-Vector Query}.
For a given dual-vector query $q$ and a query preference $\alpha$,
\oursys performs the greedy search based on the distance function (\ie Eq.\ref{eq:DVQ_dist}) to obtain the top-$k$ approximate nearest neighbors.
Specifically, \oursys maintains a candidate queue $Q_c$ and a result queue $Q_r$, where $Q_c$ contains the visited but not expanded nodes while $Q_r$ contains the expanded nodes, both ordered by their distances to the query. %
At each iteration, the first not expanded node $Q_c^\mathrm{t}$ in $Q_c$ %
is selected for expansion, i.e., its out-neighbors whose active ranges contain %
$\alpha$ are inserted into $Q_c$. The expanded node is then inserted into $Q_r$. During the search, $|Q_r|$ is bounded by $ef_{\text{search}}$. The search terminates when
$\mathrm{Dist}(q, Q_c^\mathrm{t}) >
\mathrm{Dist}(q, Q_r^\mathrm{b})$, where$Q_r^\mathrm{b}$ is the last node in $Q_r$.

\begin{figure}[t]
    \centering
    \vspace{-0.55in}
    \begin{minipage}[t]{0.48\linewidth}
        \centering
        \includegraphics[width=.85\linewidth]{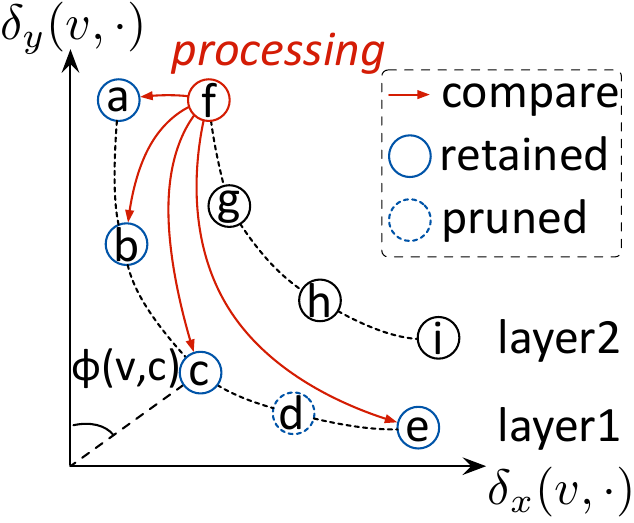}
        \vspace{-0.15in}
        \caption{Illustration of the existing AR computation algorithm~\cite{deg}.}
        \label{fig:insight_a}
    \end{minipage}
    \hfill
    \begin{minipage}[t]{0.48\linewidth}
        \centering
        
        \includegraphics[width=.83\linewidth]{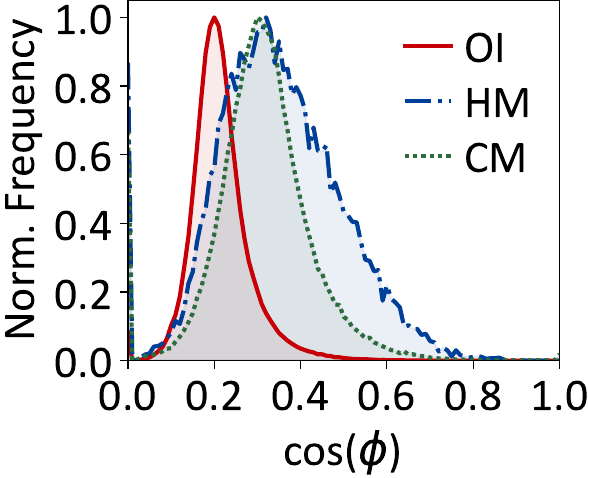}
        \vspace{-0.15in}
        \caption{
        {Distribution of retrieved nodes across different Pareto angles $\phi$ under $\alpha=0.1$.}
        }
        \label{fig:insight_b}
    \end{minipage}
\vspace{-0.15in}
\end{figure}

\vspace{-0.05in}
\section{Angle-Constrained Active Range Computation}
\label{sec:construct}

The workflow reveals that active range (AR) computation is invoked during both index construction and updates. This additional overhead substantially degrades both construction and update efficiency. To reduce this cost, we exploit the visited characteristics of Pareto-frontier-based neighbors and introduce an angle-constrained AR computation algorithm that improves computational efficiency.

\vspace{-0.1in}
\subsection{Pareto Angle}
\label{sec:pareto-angle}

The existing AR computation algorithm~\cite{deg} computes the active ranges of the candidate set $\mathcal{C}$ of a node in a nested-loop manner.
The \textbf{forward traversal order} of $\mathcal{C}$ follows a \textbf{two-level sorting scheme}: nodes are first ordered by the layer of the Pareto frontiers in ascending order and then sorted within each layer $PF$ by their distance to $v$ along a given dimension (e.g., $\delta_x(v, \cdot)$).
Specifically, when processing node $v$, \ddeg sequentially traverses its out-neighbors $(a, b, c, d, e, f, \dots)$ to compute their active ranges, following a two-level order: from lower to higher Pareto-frontier layers (i.e., layer 1 to $l$), and within each layer in ascending order of $\delta_x(v, \cdot)$, i.e., from node $a$ to $e$ 
as illustrated in Figure~\ref{fig:insight_a}.
When traversing node $f$, \ddeg compares it against all previously retained nodes, %
\ie $a-e$. 
However, this strategy performs redundant comparisons for each candidate neighbor. For $f$, it is unnecessary to compare $f$ with $e$, since $f$ is close to the $\delta_y(v,\cdot)$ axis while $e$ lies closer to the $\delta_x(v,\cdot)$ axis, making it highly unlikely that they will be simultaneously retrieved during query processing.
Before introducing our AC computation method, %
we first introduce Pareto Angle.

\begin{definition}[Pareto Angle ($\phi$)]
\label{def:pareto_angle}
Given a node $v$ and its out-neighbor $c$, the Pareto angle $\phi(v,c)$ is defined in the 2D space induced by $(\delta_x(v,c), \delta_y(v,c))$.
We define
\begin{equation}
\label{eq:phi_cmp}
\phi(v,c)
=
\arccos\left(
\frac{(\delta_x(v,c), \delta_y(v,c)) \cdot (0,1)}
{\sqrt{\delta_x(v,c)^2 + \delta_y(v,c)^2}}
\right),
\end{equation}
where $\phi(v,c)$ denotes the angle between the vector $\left(\delta_x(v,c), \delta_y(v,c)\right)$ and the positive $\delta_y(v,\cdot)$-axis. The value of $\phi(v,c)$ lies in $[0, \pi/2]$.
\end{definition}

\stitle{Observation}.
{We construct Pareto-frontier-based indexes on three real-world datasets, OI, HM, and CM (details in Table \ref{tab:dataset}), with all edges assigned an active range of $[0,1]$.}
We perform a search under query $\alpha=0.1$ and obtain the result queue $Q_r$ for each query, which contains the $ef_{{search}}$ closest retrieved nodes. We further compute the Pareto angle $\phi(v,r)$ for each retrieved node $r \in Q_r$, where $v$ is the in-neighbor that triggers the insertion of $r$ into the candidate queue during the expansion of $v$.
After the search, we count the frequency distribution of all retrieved nodes across different cosine values $\cos(\phi)$, as shown in Figure~\ref{fig:insight_b}.
From Figure~\ref{fig:insight_b}, we make the following key observation:

\etitle{Pareto-Angle Locality}.
Retrieved nodes exhibit strong locality in the Pareto angle space, where high-frequency Pareto angle regions form contiguous intervals. For instance, when $\alpha = 0.1$, nodes with higher retrieval frequency are mainly distributed in the range of $\cos(\phi) \in [0.2, 0.4]$.

\stitle{Intuition}. Based on the observation, %
it is unnecessary for each neighbor to be compared against all other neighbors when computing its active range.
Instead, comparisons should be restricted to neighbors with small differences in Pareto angle values.
This is because such neighbors are more likely to be jointly retrieved under a given preference parameter $\alpha$, and hence jointly participate in forming triangles with the target node during graph traversal.

\vspace{-0.15in}
\subsection{Active Range Computation Based on Pareto Angle Difference}
\label{sec:acprune}

Based on the above intuition, we propose an \underline{a}ngle-\underline{c}onstrained \underline{a}ctive \underline{r}ange computation algorithm, \textit{\acprune}, which leverages the Pareto angle constraint to avoid unnecessary distance computations. %
The core idea of \acprune is that, for any two candidate neighbors $e$ and $f$ of a node $v$, if both are retrieved during the search under a given $\alpha$, the absolute difference between their Pareto angles, i.e., $|\phi(v,e)-\phi(v,f)|$, should be constrained in a scope.

Specifically, when computing the active ranges of a node's candidate neighbors, \acprune employs different strategies for candidates in the first Pareto-frontier layer and those in the remaining layers, as shown in Algorithm~\ref{alg:ours_prune}.
This is because nodes in the first layer are more likely to lie on the exact Pareto frontier, and thus require more precise active range computation.

\begin{myfloat}[t]
\removelatexerror
\setstretch{0.8}
\begin{algorithm}[H]
\caption{ACAR($\mathcal{C}, R, \tau, \theta_{th}$)}
\label{alg:ours_prune}
\KwIn{Candidate set $\mathcal{C}=\{PF^1, \dots, PF^l\}$ organized in a two-level order, maximum number of edges $R$, threshold $\tau$ to prune edges with small active ranges, and angular threshold $\theta_{th} \in [0, \frac{\pi}{2}]$}
\KwOut{Neighbor set ${NS}$}

${NS} \gets \emptyset$\tcp*{Retained nodes}
\ForEach{$PF^i \in \mathcal{C}$}{
    \If(\tcp*[f]{The first layer}){$i = 1$}{
        Compute active ranges of each node in $PF^1$ in forward order;\label{line:natural_traverse}\\
        Compute active ranges of each node in $PF^1$ in reverse order;\label{line:reverse_traverse}\\

    }
    \Else(\tcp*[f]{The remaining layers}){
        \ForEach{$p \in PF^i$}{
            $ar^-_p \gets \emptyset$\tcp*{The pruning range}
            \ForEach{$w \in NS$}{
                \If{$|\phi(v,p)-\phi(v,w)| \le \theta_{th}$}{\label{line:angle_cmp}
                    compute pruning range of $p$ induced by $w$, $ar_{p}^{\prime -}$\;\label{line:pr_cmp}
                    $ar^-_p \gets ar^-_p \cup ar_{p}^{\prime -}$\;
                }
                
            }
            $ar_p \gets [0,1] \setminus ar_p^-$\tcp*{The active range}
            \If{$|ar_p| \ge \tau$}{
                $NS \gets NS \cup \{p\}$\;
            }
            \If{$|NS| \ge R$}{
                \textbf{break}\;
            }
        }
    }
}
\Return{$NS$}\;
\end{algorithm}
\vspace{-0.25in}
\end{myfloat}

\etitle{(1) The First Layer}.
In \ddeg, the first traversed node is assigned an active range of $[0,1]$ simply because the neighbor set $NS$ is initially empty, which is clearly inaccurate. 
More importantly, the active range computation for nodes in the remaining layers depends on the active ranges of nodes in the first layer, making the accuracy of the first-layer active ranges crucial.
To address this, \acprune employs a {bidirectional} two-stage scan over the first-layer candidate neighbors to compute more accurate active ranges.
Specifically, in the first-stage scan, \acprune traverses the first-layer candidate neighbors in forward order to obtain the most reliable active ranges of the final retained node in $NS$ (line~\ref{line:natural_traverse}). After completing the forward scan, we perform a reverse scan with the same pruning operations (line~\ref{line:reverse_traverse}), refining the active ranges of the first-layer neighbors.

\etitle{(2) The Remaining Layers}.
Since the proportion of candidate neighbors in the first layer is relatively small (only 4.53\% on the OI dataset), the majority of the computational cost during the AR computation is incurred by computing the active ranges of candidate neighbors in the remaining layers. 
When computing the active ranges of each neighbor, \ddeg compares it with all nodes currently retained in $NS$, which leads to substantial computational overhead. 
To address this, inspired by the observation in Section \ref{sec:pareto-angle}, \acprune leverages a Pareto angle threshold $\theta_{th}  \in [0, \frac{\pi}{2}]$ (see line~\ref{line:angle_cmp}) to skip distance computations with a subset of nodes that violate the angular constraint, thereby improving the efficiency of active range computation.
It is worth noting that the angular-constraint computation (line~\ref{line:angle_cmp}) derived from Eq.~\eqref{eq:phi_cmp} incurs much lower cost than the pruning range computation (line~\ref{line:pr_cmp}). This is because the distance values required for Pareto-angle computation, i.e., $\delta_x$ and $\delta_y$, are already maintained during candidate construction and can be accessed directly without additional distance evaluations.

\begin{figure}[tbp]
\centering
\vspace{-0.4in}

\begin{subfigure}[t]{0.45\linewidth}
    \centering
    \includegraphics[width=.85\linewidth]{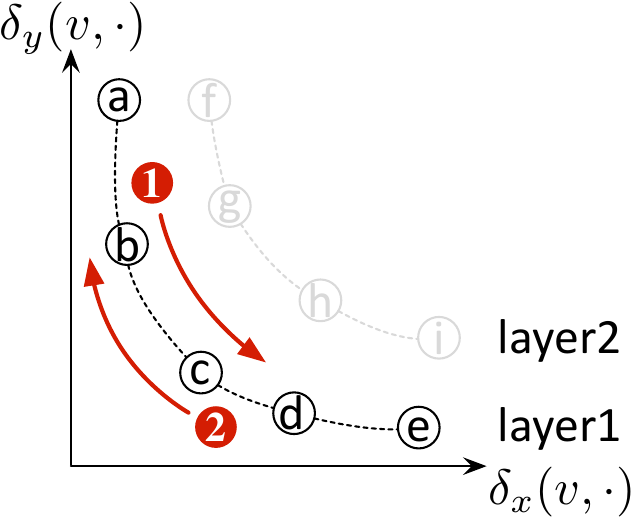}
    \vspace{-0.08in}
    \caption{The first layer.}
    \label{fig:ours_prune_1}
\end{subfigure}
\hfill
\begin{subfigure}[t]{0.45\linewidth}
    \centering
    \includegraphics[width=.85\linewidth]{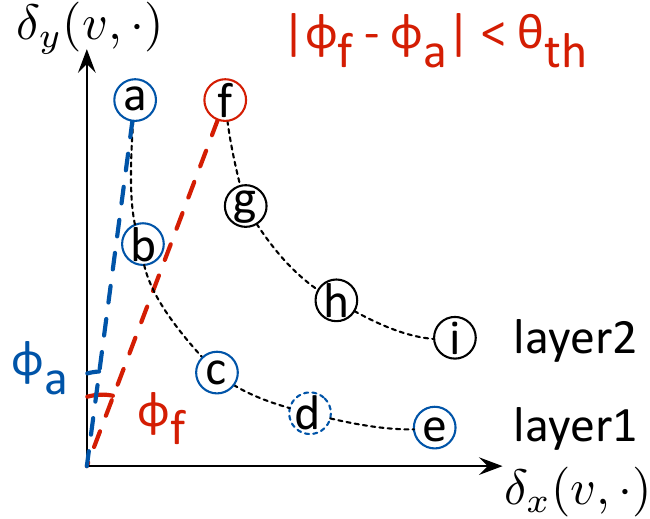}
    \vspace{-0.08in}
    \caption{The remaining layers.}
    \label{fig:ours_prune_2}
\end{subfigure}

\vspace{-0.15in}
\caption{Illustration of \acprune.}
\label{fig:ours_prune}
\vspace{-0.2in}
\end{figure}

\begin{example}
Figure~\ref{fig:ours_prune} illustrates how to compute the active ranges of candidate neighbors for node $v$.
{For the candidate neighbors in the first layer}, \acprune traverses them in forward order (e.g., the $x$-dimension), proceeding from node $a$ to node $e$, as shown in Figure~\ref{fig:ours_prune_1}\circlednum{1}. 
Through this traversal, \acprune computes the active ranges of node $e$ and inserts it into the neighbor set $NS$. 
Subsequently, \acprune traverses the remaining candidates of the first layer in reverse order to compute their active ranges, proceeding from node $d$ back to node $a$, as illustrated in Figure~\ref{fig:ours_prune_1}\circlednum{2}.
{For candidate neighbors in the remaining layers}, we take node $f$ as an example, as shown in Figure~\ref{fig:ours_prune_2}.
At this point, nodes $a$, $b$, $c$, and $e$ remain in the current neighbor set $NS$, while node $d$ has been pruned.
\acprune compares node $f$ with all nodes in $NS$ that satisfy the angle constraint.
For instance, for node $a$, \acprune computes $|\phi(v,a)-\phi(v,f)|$.
If $|\phi(v,a)-\phi(v,f)| \le \theta_{th}$, \acprune computes the pruning range induced by $a$, otherwise, the computation is skipped.
\end{example}

\vspace{-0.1in}
\subsection{Discussion}
\label{subsec:complexity_prune}
\vspace{-0.05in}

\stitle{Applicability}. {\acprune is specifically designed for DVQ, where each object is associated with exactly two attribute vectors. Extending \acprune to settings with more than two attribute vectors introduces substantially greater complexity and is far from a trivial task.}

\vspace{-0.05in}
\stitle{Benefits of \acprune}.
{\acprune introduces an angle constraint to reduce unnecessary distance computations.
Given a target node $v$ with candidate neighbor set $\mathcal{C}$, when processing the $i$-th candidate neighbor $p_i$, \acprune compares $p_i$ against a subset $NS_i' \subseteq NS_i$ of the currently retained nodes that satisfy the angle constraint, where $NS_i$ denotes the retained node set at iteration $i$ and $t \le |\mathcal{C}|$ is the total number of iterations.
Therefore, the total number of distance computations is $\sum_{i=1}^{t}|NS_i'|$, where $|NS_i'| \le |NS_i|$.
The reduction in distance computations achieved by \acprune over the baseline depends on the distribution of Pareto angles. When Pareto angles are highly concentrated, \acprune may degenerate to the baseline, whereas a more uniform Pareto-angle distribution allows \acprune to fully exploit its acceleration potential.
}

\section{Connectivity-Loss-Driven Incremental Pareto Frontier Repair}
\label{sec:deletion}
As discussed in Section~\ref{sec:bg_update}, %
applying single-vector repair methods for DVQ indices that incorporate Pareto frontiers is unable to simultaneously ensure update efficiency and retrieval effectiveness.
To address this issue, we introduce a \emph{connectivity-loss-driven adaptive repair algorithm}
to maintain graph connectivity under deletions while avoiding unnecessary repair operations. %

\vspace{-0.1in}
\subsection{Observation \& Intuition}
\label{sec:cl_def}

For each affected node, reinsertion-based repair (RBR) methods perform Greedy Pareto Frontier Search from entry nodes to reconstruct its candidate neighbors. However, reinsertion incurs substantial redundant computations, resulting in inefficient updates.

\begin{figure}[t]
    \centering
    \vspace{-0.4in}
    \begin{subfigure}{.48\linewidth}
        \centering
        \includegraphics[width=.9\linewidth]{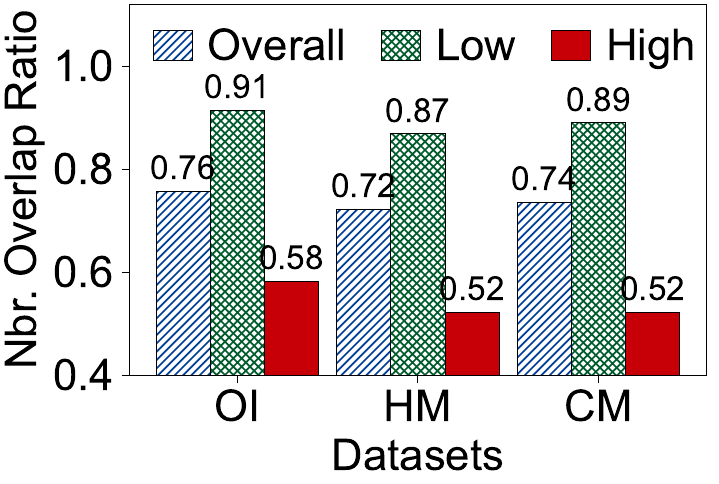}
        \vspace{-0.05in}
        \caption{Neighbor Similarity}\label{fig:proportion}
    \end{subfigure}
    \hfill
    \begin{subfigure}{.48\linewidth}
        \centering
        \includegraphics[width=0.9\linewidth]{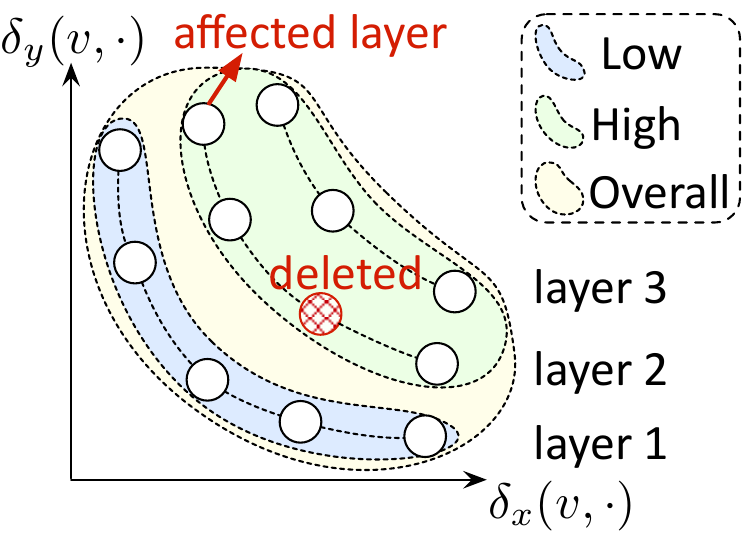}
        \vspace{-0.05in}
        \caption{Example of an affected node}\label{fig:frontiers_exp}
    \end{subfigure}
    \vspace{-0.15in}
    \caption{Neighbor similarity of affected nodes before and after reinsertion-based repair.}
    \label{fig:neighbor_similarity}
    \vspace{-0.2in}
\end{figure}

\stitle{Observation}.
To investigate the feasibility of incremental repair in Pareto-frontier-based indices under deletions, we measure the out-neighbor overlap ratio of affected nodes on three real-world datasets, as shown in Figure~\ref{fig:proportion}.
{
Specifically, we compute the overall overlap ratio of the out-neighbors for each affected node $v$, i.e.,
$\frac{|N'_{{out}}(v) \cap N_{{out}}(v)|}{|N_{{out}}(v)|}$,
where $N_{{out}}(v)$ and $N'_{{out}}(v)$ denote the out-neighbor sets before and after reinsertion, respectively.
Furthermore, we analyze the overlap ratio across different layers of Pareto frontiers.
Here, affected layers refer to layers that contain deleted nodes (\ie layer~2). Low layers are defined as all layers below the lowest affected layer (\ie layer~1), while high layers refer to the affected layers and all layers above them (\ie layer~2 and layer~3), as illustrated in Figure~\ref{fig:frontiers_exp}.
}
We observe that 1) \textit{the overall overlap ratio is at least \textbf{72\%}, indicating that most original neighbors remain effective after repair}, and 2) \textit{87\% of neighbors in low layers are preserved, whereas only 52\% in high layers are retained}.

\stitle{Intuition}.
These observations provide strong support for incremental repair, which avoids the costly global search over all neighbors required by the reinsertion-based method.
Observation 1) suggests that most original neighbors of an affected node still belong to its approximate Pareto frontiers, enabling incremental repair during deletions.
Furthermore, Observation 2) suggests that neighbor updates should be performed selectively across layers. Specifically, neighbors in low layers remain unchanged, while those in high layers need to be updated.

\vspace{-0.1in}
\subsection{Pareto Frontier Repair Based on Connectivity Loss}
\label{sec:ca-repair}

We further analyze the distribution of deleted neighbors among affected nodes under a 1\% deletion batch on the OpenImage index, with a maximum neighbor size of $R=40$.
The results show that the number of deleted neighbors varies significantly across affected nodes (\ie in-neighbors of deleted nodes), ranging from 1 to 20, indicating that the degree of connectivity loss induced by deletions differs substantially among affected nodes.
We next provide a formal definition of connectivity loss.

\begin{definition}[Connectivity Loss ($\mathrm{CL}$)]
\label{def:cl}
The connectivity loss of a node $v$ measures the number of its lost out-neighbors, and its degree is defined as
\begin{equation}
\mathrm{CL}(v) \triangleq \frac{R - |N_{out}(v)|}{R},
\end{equation}
where $R$ denotes the maximum number of neighbors and $N_{out}(v)$ denotes the current out-neighbor set of $v$. It holds that $\mathrm{CL}(v) \in [0,1]$.
\end{definition}

Ideally, each node in the index maintains $R$ out-neighbors, resulting in no connectivity loss ($\mathrm{CL}=0$). 
In practice, due to imperfect construction or deletions, nodes may suffer from connectivity loss. 
To this end, we categorize the degree of connectivity loss into three levels, mild, moderate, and severe. 
Specifically, for a node $v$, mild connectivity loss corresponds to $\mathrm{CL}(v) \in [0, {CL}_1]$, moderate connectivity loss corresponds to $\mathrm{CL}(v) \in ({CL}_1, {CL}_2]$, and severe connectivity loss corresponds to $\mathrm{CL}(v) \in ({CL}_2, 1]$, where $CL_1$ and $CL_2$ are the connectivity loss thresholds.

\stitle{Connectivity-Loss-Driven Adaptive Repair}.
Based on the above analysis, %
we design a \underline{c}onnectivity-loss-driven \underline{a}daptive \underline{P}areto \underline{f}rontier \underline{r}epair algorithm, \pfrepair, to restore the connectivity of the graph index.
The core idea of \pfrepair is to determine the repair intensity for affected nodes according to the degree of connectivity loss.
The detailed procedure of \pfrepair is shown in Algorithm~\ref{alg:CA-PFR}.

Specifically, when a set of nodes $\mathcal{D}$ is deleted, \pfrepair repairs the affected nodes $V_{affected}$, i.e., the in-neighbors of the deleted nodes, according to their degree of connectivity loss. 
For each affected node $v \in V_{{affected}}$, \pfrepair first
organizes its out-neighbors into multi-layer Pareto frontiers $\mathcal{C}=\{PF^1_v, \dots, PF^l_v\}$ (line~\ref{line:getL}), and then obtains its active out-neighbors $Out \gets N_{{out}}(v) \setminus \mathcal{D}$ (line~\ref{line:getO}) and active in-neighbors $In \gets N_{{in}}(v) \setminus \mathcal{D}$ (line~\ref{line:getI}).
Next, the degree of connectivity loss for node $v$, $\mathrm{CL}(v)$, is computed (line~\ref{line:getDelta}) according to Definition~\ref{def:cl}.
Based on $\mathrm{CL}(v)$, \pfrepair adopts historical-Pareto-frontier repair, affected-layer repair, and multi-layer repair to handle nodes with mild, moderate, and severe connectivity loss, respectively.

\begin{myfloat}[t]
\vspace{-0.5in}
\removelatexerror
\setstretch{0.8}
\begin{algorithm}[H]
\caption{CA-PFR}
\label{alg:CA-PFR}
\KwIn{DVQ graph $G=(V,E)$, deletion set $\mathcal{D}$, affected nodes $V_{{affected}}$, out-degree bound $R$,
thresholds ${CL}_1,{CL}_2$, threshold $\tau$ and $\theta_{th}$}
\KwOut{Repaired graph on nodes $V' = V \setminus \mathcal{D}$}

\BlankLine
\ForEach{$v \in V_{{affected}}$}{
    $\mathcal{C} \gets \{PF^1_v, \dots, PF^l_v\}$\tcp*{Candidate set}\label{line:getL}
    $\mathcal{C}' \gets \emptyset$\tcp*{Repaired candidate set}
    $Out \gets N_{{out}}(v) \setminus \mathcal{D}$\;\label{line:getO}
    $In \gets N_{{in}}(v) \setminus \mathcal{D}$\; \label{line:getI}
    
    $\mathrm{CL}(v) \gets \frac{R - |Out|}{R}$\; \label{line:getDelta}
    
    \uIf(\tcp*[f]{Mild}){$0 < \mathrm{CL}(v) \le {CL}_1$}{
        $\mathcal{C}' \gets Out \cup In$\; \label{line:mildC}
    }
    \uElseIf(\tcp*[f]{Moderate}){${CL}_1 < \mathrm{CL}(v) \le {CL}_2$}{
        $\mathcal{C}_{affected} \gets \{ PF_v^i \in \mathcal{C} \mid PF_v^i \cap \mathcal{D} \neq \emptyset \}$\;\label{line:affectedLD}
        $Out^{2} \gets \bigcup_{PF \in \mathcal{C}_{affected}} \ \bigcup_{p \in PF} \left( N_{{out}}(p) \setminus \mathcal{D} \right)$\; \label{line:affectedO2}
        $\mathcal{C}' \gets Out \cup In \cup Out^{2}$\; \label{line:affectedC}
    }
    \Else(\tcp*[f]{Severe}){
        $\ell_{\min} \gets \arg\min_{i} \; i \;\text{s.t.}\; PF^i_v \cap \mathcal{D} \neq \emptyset$\; \label{line:lmin}
        $\mathcal{C}_{high} \gets \{ PF_v^i \in \mathcal{C} \mid i \ge \ell_{\min} \}$\; \label{line:highLD}
        $Out^{2} \gets \bigcup_{PF \in \mathcal{C}_{high}} \ \bigcup_{p \in PF} \left( N_{{out}}(p) \setminus \mathcal{D} \right)$\; \label{line:highO2}
        $\mathcal{C}' \gets Out \cup In \cup Out^{2}$\; \label{line:highC}
    }
    
    $N_{{out}}(v) \gets \textsc{\acprune}(\mathcal{C}', R,\tau,\theta_{th})$\; \label{line:pruneNew}
}
\end{algorithm}
\vspace{-0.15in}
\end{myfloat}

\etitle{Historical-Pareto-Frontier Repair}.
The candidate set $\mathcal{C}'$ consists of $v$'s active out-neighbors $Out$ and active in-neighbors $In$ (line~\ref{line:mildC}). 
Due to the low connectivity loss, this strategy leverages the in-neighbors of affected nodes to efficiently reduce connectivity degradation with minimal overhead.
In particular, it avoids any distance computations during candidate selection, thereby significantly reducing the overall repair cost. 
The key idea is that, due to the reverse edges introduced during index construction, the in-neighbors of a node have previously served as its candidate out-neighbors.

\etitle{Affected-Layer Repair}.
The candidate set $\mathcal{C}'$ consists of $Out$, $In$, and an additional expanded set $Out^{2}$ derived from the affected layers of Pareto frontiers (i.e., layer 2 in Figure~\ref{fig:frontiers_exp}).
Specifically, \pfrepair first locates the affected layers in $\mathcal{C}$ that contain deleted nodes, denoted by $\mathcal{C}_{affected}$ (line~\ref{line:affectedLD}).
It then expands the nodes in these affected layers to form the expanded set $Out^{2}$ (line~\ref{line:affectedO2}).
This expansion strategy is motivated by the observation that neighbors in layers below the affected layers are rarely updated during repair.
By expanding nodes in these affected layers, \pfrepair can effectively restore local connectivity while avoiding unnecessary exploration.

\etitle{Multi-Layer Repair}.
The candidate set $\mathcal{C}'$ consists of $Out$, $In$ and an expanded set $Out^{2}$ derived from all high layers (i.e., the "High" region in Figure~\ref{fig:frontiers_exp}).
Specifically, \pfrepair first locates the lowest affected layer containing deleted nodes and denotes its layer ID as $\ell_{\min}$ (line~\ref{line:lmin}).
It then collects all layers of Pareto frontiers whose layer IDs are not smaller than $\ell_{\min}$, denoted by $\mathcal{C}_{high}$ (line~\ref{line:highLD}), and gathers the active out-neighbors of nodes in these layers to form $Out^{2}$ (line~\ref{line:highO2}).
This expansion strategy is designed not only to address significant connectivity loss caused by deletions, but also to mitigate connectivity deficiencies inherited from the index construction phase, thereby further improving overall graph connectivity.
For example, in the LAION dataset, approximately $2.77\%$ of nodes in the initial index exhibit connectivity loss above ${CL}_2$ before deletions. After 30 batches of deletions, the graph connectivity does not degrade. Instead, the number of nodes with connectivity loss above ${CL}_2$ decreases by $3.82\times$.

Finally, the repaired candidate set $\mathcal{C}'$ of the affected node $v$ is refined to obtain the final out-neighbors $N_{{out}}(v)$ (line~\ref{line:pruneNew}), thereby completing the structural repair of node $v$ in the index.

\begin{figure}[tbp]
     \vspace{-0.5in}
    \centering
    \includegraphics[width=.99\linewidth]{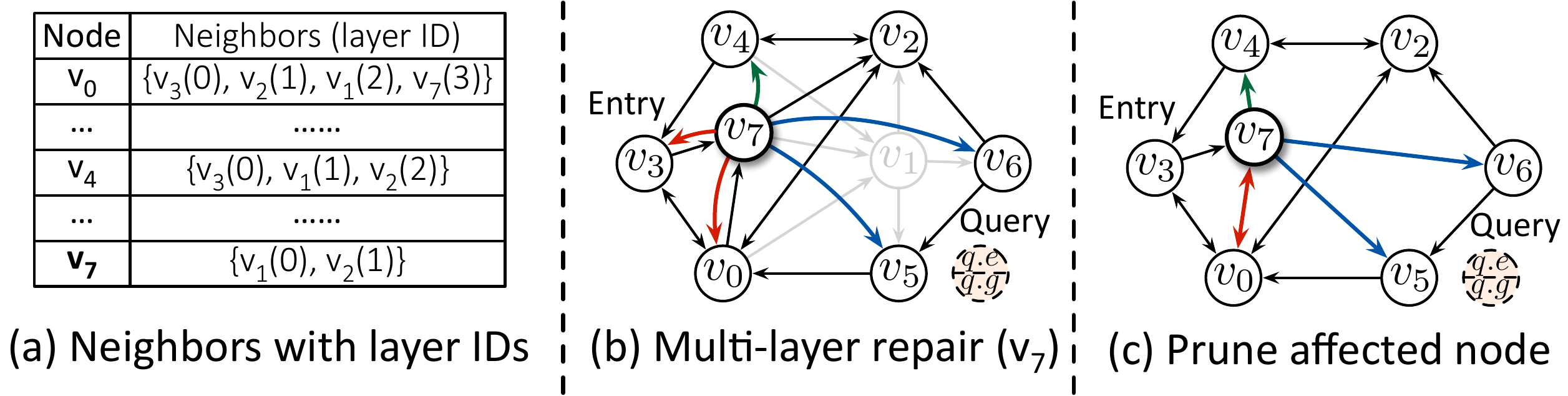}
    \vspace{-0.15in}
    \caption{An illustrative example of multi-layer repair.}
    \label{fig:example_repair}
     \vspace{-0.3in}
\end{figure}

\begin{example}
To better illustrate how \pfrepair adaptively repairs the in-neighbors affected by a deleted node, we focus on the case of severe connectivity loss, corresponding to node $v_7$, as shown in Figure~\ref{fig:example_repair}.
For the affected node $v_7$, \pfrepair first computes its connectivity loss $\mathrm{CL}(v_7) = 3/4$. Since $\mathrm{CL}(v_7)$ exceeds ${CL}_2$ (\ie 1/2), \pfrepair classifies $v_7$ as exhibiting severe connectivity loss and thus triggers the multi-layer repair strategy.
Specifically, \pfrepair employs Multi-Layer Repair to construct the final candidate set, i.e., ${v_0, v_2, v_3, v_4, v_5, v_6}$.
Finally, node $v_7$ connects to $v_0$, $v_4$, $v_5$ and $v_6$ after the AR computation, as shown in Figure~\ref{fig:example_repair}c.
Notably, this strategy compensates for inherent connectivity deficiencies in the initial index.
For example, in the initial graph index (see Figure~\ref{fig:bg_index}a), reaching $v_4$ from $v_3$ follows the path $v_3 \rightarrow v_7 \rightarrow v_2 \rightarrow v_4$ (3 hops).
In contrast, after multi-layer repair (see Figure~\ref{fig:example_repair}c), $v_3$ can reach $v_4$ via $v_7$ in only 2 hops, thereby reducing search latency.
\end{example}

\stitle{Asynchronous Reverse Graph Maintenance}.
During deletion, \oursys requires in-neighbor information to locate and repair affected nodes. Since the graph-based index stores only out-neighbors, obtaining in-neighbors would otherwise require scanning the entire graph topology $G_{out}$. To avoid this overhead, \oursys maintains a lightweight reverse graph topology $G_{in}$, where for each node $v$,
$N^{G_{in}}_{{out}}(v)=N^{G_{out}}_{{in}}(v)$. Unlike $G_{out}$, whose edges store dual-vector distances and active ranges, $G_{in}$ stores only in-neighbor IDs, accounting for merely 2.34\% of the combined size of $G_{out}$ and the dual-vector data in the HM dataset. Since batch updates modify edges in $G_{out}$, \oursys asynchronously maintains $G_{in}$ in the background to minimize synchronization overhead.

\stitle{{Distance Storage and Memory Overhead}}{.}
{
In \oursys, we store only the distances between each node and its neighbors with respect to the two attribute vectors as edge metadata in the adjacency list, which are reused during historical-Pareto-frontier repair and active range computation. When an edge is accessed, its distance values can be read in the same memory access without an additional lookup. During index updates, these distances are inserted or removed together with the edge metadata.}
{Despite the modest additional storage overhead
, this design improves update throughput
and is therefore practically meaningful.}
{For fairness, all baselines also store dual-vector distances in Section~6.
}

\vspace{-0.1in}
\subsection{Computational Overhead Analysis}
\label{subsec:analysis_cmp_repair}

As the complexity of AR computation has been discussed in Section~\ref{subsec:complexity_prune}, we focus on the complexity of candidate repair in \pfrepair.
Given a deleted node, \pfrepair adaptively augments the candidate set $\mathcal{C}$ for each affected node $v \in V_{{affected}}$, yielding a repaired candidate set $\mathcal{C}'$.
Therefore, the computational complexity for repairing an affected node $v$ is $O(|\Delta \mathcal{C}|)$, where $\Delta \mathcal{C} = \mathcal{C}' \setminus \mathcal{C}$ denotes the newly introduced candidate neighbors of $v$.
Next, we analyze the size of $|\Delta \mathcal{C}|$ under the following three repair strategies of \pfrepair.

\etitle{Historical-Pareto-Frontier Repair (Mild)}.
For each affected node $v \in V_{{affected}}$, this strategy augments the original active out-neighbors in $\mathcal{C}$ with the active in-neighbors of $v$, resulting in $|\Delta \mathcal{C}| \approx |N_{in}(v)|$.
Since the distances $\delta_x(v, v')$ and $\delta_y(v, v')$ are already stored in the corresponding in-neighbors $v' \in N_{in}(v)$, the candidate set augmentation incurs no additional distance computations, i.e., $T_{{mild}} = 0$.

\etitle{Affected-Layer Repair (Moderate)}.
In addition to the active in-neighbors $N_{in}(v)$, this repair strategy further adds the active out-neighbors of nodes in the affected layers to $\mathcal{C}'$, yielding $|\Delta \mathcal{C}| \approx |N_{in}(v)| + n \cdot R$, where $n$ is the average number of nodes per layer, $R$ denotes the maximum out-degree, and $n \le R$.
Thus, the complexity for repairing each affected node $v$ is:
$T_{{moderate}} \approx O\left(n \cdot R\right)$.

\etitle{Multi-Layer Repair (Severe)}.
This repair strategy not only includes the in-neighbors and the original active out-neighbors, but also incorporates all active out-neighbors of nodes in the high layers into $\mathcal{C}'$, resulting in $|\Delta \mathcal{C}| \approx |N_{in}(v)|+ h \cdot n \cdot R$, where $h$ is the number of the high layers, and $h \cdot n \le R$.
Therefore, the computational complexity of repairing each affected node $v$ is:
$T_{{severe}} \approx O\left(h \cdot n \cdot R\right)$.

Let $P_1$, $P_2$, and $P_3$ denote the probabilities of triggering the three repair strategies, respectively, satisfying $P_1 + P_2 + P_3 = 1$.
As a result, the computational complexity of \pfrepair is:
$T_{{\pfrepair}}=P_1 \cdot T_{{mild}}+ P_2\cdot T_{{moderate}} + P_3 \cdot T_{{severe}}$.
In practice, since severe repair is triggered infrequently, with probabilities satisfying $P_1 > P_2 \gg P_3$, the expected complexity can be simplified as:
$
\mathbb{E}[T_{{\pfrepair}}]
=P_2' \cdot T_{{moderate}}$,
where
$P_2' = \frac{P_2}{P_1 + P_2}$.

\section{Evaluation}
\label{sec:expr}

\begin{figure*}[thp]
    \centering
    \vspace{-0.5in}
    \begin{subfigure}[t]{0.35\linewidth}
        \centering
        \includegraphics[width=\linewidth]{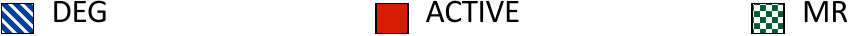}
    \end{subfigure}

     \vspace{-0.04in}

    \begin{subfigure}[t]{0.19\linewidth}
        \centering
        \includegraphics[width=\linewidth]{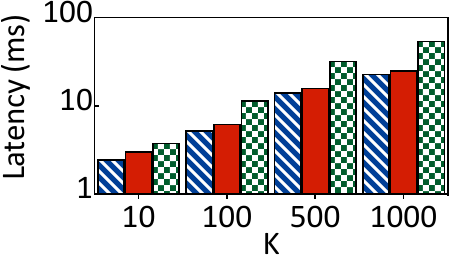}
        \vspace{-0.25in}
        \caption{OI}
    \end{subfigure}
    \hfill
    \begin{subfigure}[t]{0.19\linewidth}
        \centering
        \includegraphics[width=\linewidth]{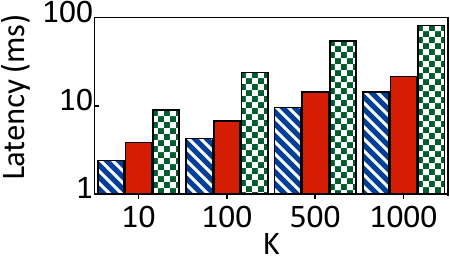}
        \vspace{-0.25in}
        \caption{HM}
    \end{subfigure}
    \hfill
    \begin{subfigure}[t]{0.19\linewidth}
        \centering
        \includegraphics[width=\linewidth]{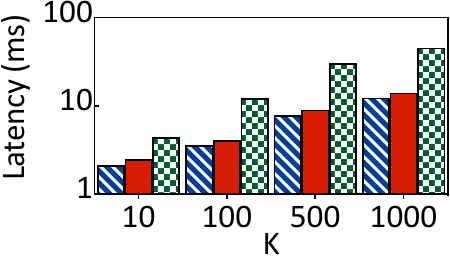}
        \vspace{-0.25in}
        \caption{CM}
    \end{subfigure}
    \hfill
    \begin{subfigure}[t]{0.19\linewidth}
        \centering
        \includegraphics[width=\linewidth]{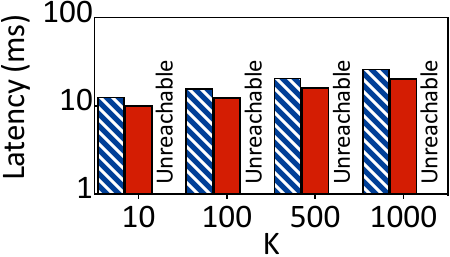}
        \vspace{-0.25in}
        \caption{LN}
    \end{subfigure}
    \hfill
    \begin{subfigure}[t]{0.195\linewidth}
        \centering
        \includegraphics[width=\linewidth]{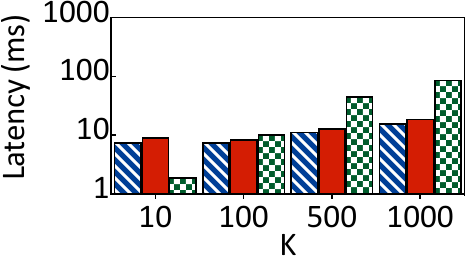}
        \vspace{-0.25in}
        \caption{CW}
    \end{subfigure}

    \vspace{-0.15in}
    \caption{{Query latency comparison at a fixed recall$= 0.95$ when varying $k$.}}
    \label{fig:diffk}
    \vspace{-0.15in}
\end{figure*}

\begin{figure}[t]
    \centering

        \begin{minipage}[t]{0.45\linewidth}
        \centering
        \includegraphics[width=\linewidth]{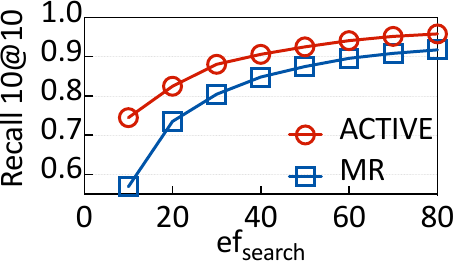}
        \vspace{-0.25in}
        \caption{{Recall comparison under different candidate budgets on OI.}}
        \label{fig:diff_ef}
    \end{minipage}
    \hfill
\begin{minipage}[t]{0.44\linewidth}
        \centering
        
        \includegraphics[width=\linewidth]{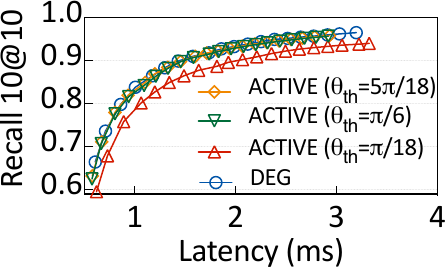}
        \vspace{-0.25in}
        \caption{{Search performance with varying Pareto angle thresholds $\theta_{th}$.}}
        \label{fig:diff_theta_hm}
    \end{minipage}
\vspace{-0.15in}
\end{figure}

\begin{figure}[tbp]
    \centering
    \begin{subfigure}[t]{0.7\linewidth}
        \centering
        \includegraphics[width=\linewidth]{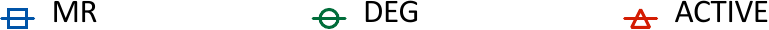}
    \end{subfigure}
    \vspace{-0.1in}
    \begin{subfigure}[t]{0.32\linewidth}
        \centering
        \includegraphics[width=\linewidth]{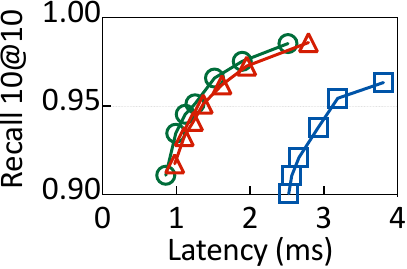}
        \vspace{-0.2in}
        \caption{$\alpha \sim \mathrm{Beta}(6.4, 13.6)$}
    \end{subfigure}
    \hfill
    \begin{subfigure}[t]{0.32\linewidth}
        \centering
        \includegraphics[width=\linewidth]{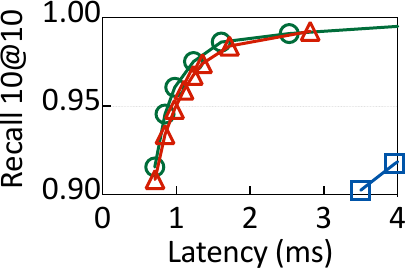}
        \vspace{-0.2in}
        \caption{$\alpha \sim \mathrm{Beta}(10, 10)$}
        \label{fig:diff_alpha_0.5}
    \end{subfigure}
    \hfill
    \begin{subfigure}[t]{0.32\linewidth}
        \centering
        \includegraphics[width=\linewidth]{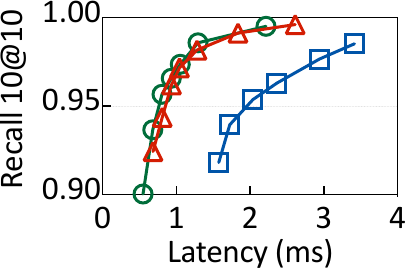}
        \vspace{-0.2in}
        \caption{$\alpha \sim \mathrm{Beta}(13.6, 6.4)$}
    \end{subfigure}
    \vspace{-0.1in}
    \caption{{Comparison of the %
    search accuracy and latency under non-uniform $\alpha$ distributions.}}
    \label{fig:diff_alpha}
    \vspace{-0.25in}
\end{figure}

\subsection{Experimental Setup}
\label{subsec:expr_setup}

\stitle{Evaluation Platform}.
All experiments are conducted on a server with an Intel(R) CPU, %
320\,GB %
memory. %
The system runs Ubuntu 22.04 LTS with GCC version 11.4.0.

\begin{table}[tbp]
\centering
\vspace{-0.3in}
\caption{Dataset statistics. $D_x$ and $D_y$ denote the vector dimensions of the $x$- and $y$-attributes, respectively. Each dataset contains 1,000 query vectors, {except Caselaw (88.3K)}.}
\label{tab:dataset}
\vspace{-0.1in}

\begin{tabular}{lccrl}
\toprule
Dataset & $D_x$ & $D_y$ & \#Vectors & Contents \\
\midrule
OpenImage (OI) & 768  & 768 & 507,444    & Text-Image \\
Howto100M (HM) & 1024 & 768 & 1,238,875  & Text-Video \\
CC3M (CM)      & 768  & 768 & 3,131,153  & Text-Image \\
LAION (LN)     & 512  & 512 & 10,642,155 & Text-Image \\
{Caselaw (CW)}   & {1,536} & {2}   & {7,288,481}    & {Text-Coord.} \\
\bottomrule
\end{tabular}
\vspace{-0.2in}
\end{table}

\stitle{Datasets}.
We evaluate our system on five publicly available real-world datasets, namely OpenImage (OI)~\cite{deg}, Howto100M (HM)~\cite{deg}, CC3M (CM)~\cite{deg}, LAION (LN)~\cite{laion} and {Caselaw (CW)}~\cite{caselaw}, as summarized in Table~\ref{tab:dataset}.
{For CW, we construct its DVQ version using the original single text embeddings and associated court location metadata.}
{Note that \oursys is not restricted to the attribute-type combinations in Table~\ref{tab:dataset}, but generalizes to arbitrary dual-vector data types. %
}

\stitle{Compared Methods}.
We comprehensively evaluate \oursys against baselines to validate its search and update performance.

\etitle{Static Search}.
We select multi-index retrieval (MR)~\cite{multi-recall,milvus} and DEG~\cite{deg} as baselines for static search, where MR builds a separate single-vector index (i.e., an HNSW index) for each of the two attribute vectors.
The difference between \oursys and \ddeg in construction lies in their active range computation methods.

\etitle{Dynamic Updates}.
Since no dedicated methods exist for dynamic DVQ index updates, we use RBR~\cite{IPGM} and NRBR~\cite{FreshDiskANN} for deletions and DEG's incremental construction strategy for insertions.
These baselines employ existing AR computation algorithms~\cite{deg}.

\stitle{{Parameter Settings}}{.}
{During index construction, we set $ef_{\mathrm{construct}}=200$ and $R$ (or $M$) = $40$ for \oursys, DEG and MR.
On LN, $ef_{\mathrm{construct}}$ is increased to $400$, because setting them to $200$ yields an average out-degree of only about $R/2$ due to the data distribution. The other parameters for DEG and MR follow those recommended in their respective papers.
The Pareto angle threshold of \oursys is set to $\theta_{th}=\pi/18$.
During updates, we set $ef_{\mathrm{update}}=100$ for insertions in \oursys and DEG, and for reinsertions in RBR, and} {use $ef_{\mathrm{search}}=100$ for search.}
{NRBR uses the same $R=40$. The connectivity-loss thresholds of \oursys are $\mathrm{CL}_1=1/3$ and $\mathrm{CL}_2=1/2$. Unless otherwise specified, all updates use $32$ threads, and each query $\alpha$ is randomly sampled from $[0,1]$.}

\stitle{Workloads}. 
By default, we construct a DVQ index on 100\% of each dataset for search and deletion experiments,
and another index on 50\% of each dataset for insertion experiments.
For deletions, we use the 100\% index and remove 1\% of the vectors in each of 10 batches, resulting in a total of 10\% deletions. For insertions, we use the 50\% index and insert 1\% of the vectors in each of 10 batches, resulting in a total of 10\% insertions.

\stitle{Metrics}.
We measure search accuracy using Recall@$k$ ($k=10$ by default) and search efficiency using average query latency, while update (i.e., deletions and insertions) performance is evaluated using throughput (deletions/insertions per second). %

\begin{table}[t]
\centering
\vspace{-0.3in}
\caption{Comparison of active range computation time (in seconds) across different datasets.}
\label{tab:construct}
\vspace{-0.1in}

\begin{tabular}{lccccc}
\toprule
Method & OI  & HM  & CM  & LN & CW \\
\midrule
DEG     & 1032.47 & 2476.73 & 3903.47 & 9221.83 & 20346.90 \\
\oursys & 732.33 & 1351.25  & 2539.33 & 6902.11 & 16220.81\\
\bottomrule
\end{tabular}

\vspace{-0.2in}
\end{table}

\vspace{-0.1in}
\subsection{Static Search}
\label{subsec:expr_static}

\begin{figure*}[t]
    \centering
    \vspace{-0.5in}
    \begin{subfigure}[t]{0.35\linewidth}
        \centering
        \includegraphics[width=\linewidth]{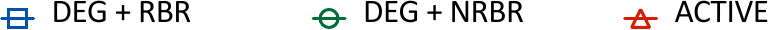}
        \vspace{-0.2in}
    \end{subfigure}

    \begin{subfigure}[t]{0.19\linewidth}
        \centering
        \includegraphics[width=\linewidth]{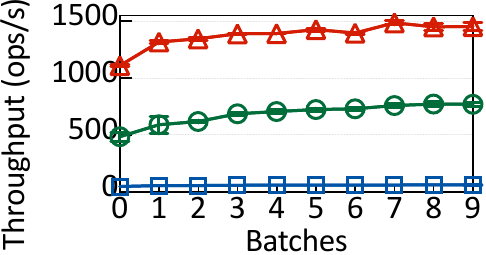}
        \vspace{-0.2in}
        \caption{OI}
    \end{subfigure}
    \hfill
    \begin{subfigure}[t]{0.19\linewidth}
        \centering
        \includegraphics[width=\linewidth]{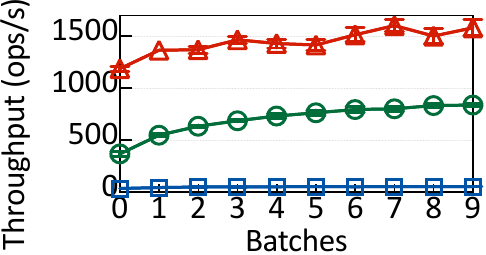}
        \vspace{-0.2in}
        \caption{HM}
    \end{subfigure}
    \hfill
    \begin{subfigure}[t]{0.19\linewidth}
        \centering
        \includegraphics[width=\linewidth]{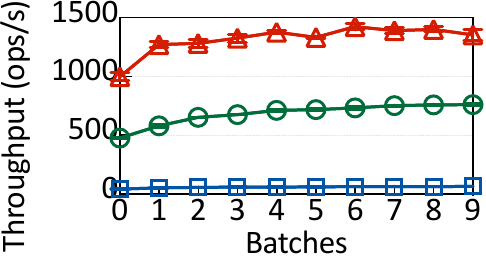}
        \vspace{-0.2in}
        \caption{CM}
    \end{subfigure}
    \hfill
    \begin{subfigure}[t]{0.19\linewidth}
        \centering
        \includegraphics[width=\linewidth]{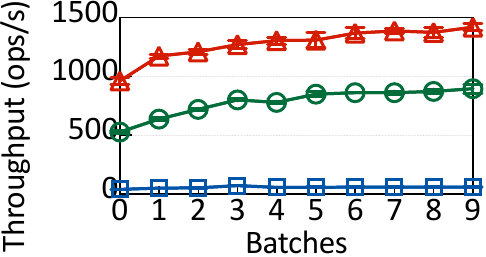}
        \vspace{-0.2in}
        \caption{LN}
    \end{subfigure}
    \hfill
    \begin{subfigure}[t]{0.19\linewidth}
        \centering
        \includegraphics[width=\linewidth]{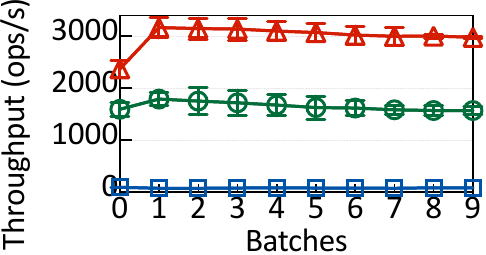}
        \vspace{-0.2in}
        \caption{CW}
    \end{subfigure}
    \vspace{-0.15in}
    \caption{{Throughput comparison with standard deviation over three runs under deletions.}}
    \label{fig:delete_ops}
     \vspace{-0.15in}
\end{figure*}

\begin{figure*}[tbp]
    \centering

    \begin{subfigure}[t]{0.35\linewidth}
        \centering
        \includegraphics[width=\linewidth]{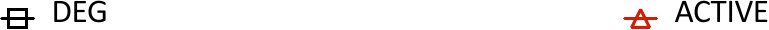}
        \vspace{-0.2in}
    \end{subfigure}

    \begin{subfigure}[t]{0.19\linewidth}
        \centering
        \includegraphics[width=\linewidth]{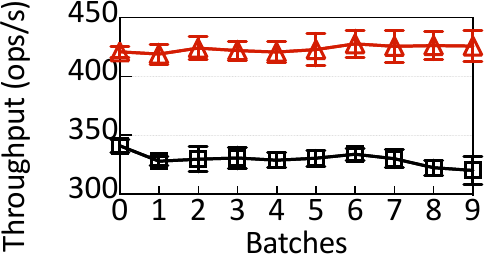}
        \vspace{-0.2in}
        \caption{OI}
    \end{subfigure}
    \hfill
    \begin{subfigure}[t]{0.19\linewidth}
        \centering
        \includegraphics[width=\linewidth]{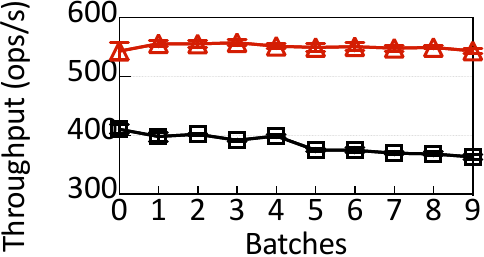}
        \vspace{-0.2in}
        \caption{HM}
    \end{subfigure}
    \hfill
    \begin{subfigure}[t]{0.19\linewidth}
        \centering
        \includegraphics[width=\linewidth]{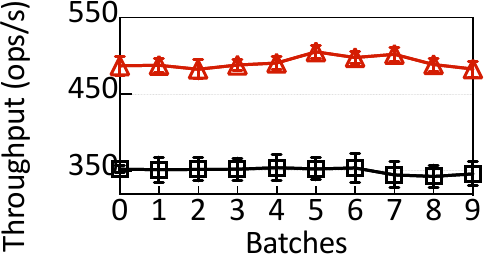}
        \vspace{-0.2in}
        \caption{CM}
    \end{subfigure}
    \hfill
    \begin{subfigure}[t]{0.19\linewidth}
        \centering
        \includegraphics[width=\linewidth]{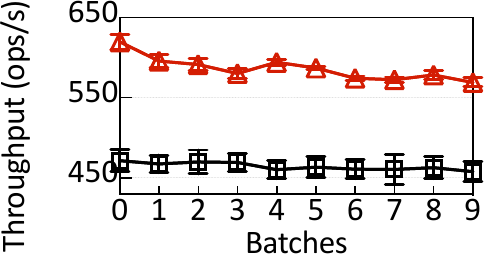}
        \vspace{-0.2in}
        \caption{LN}
    \end{subfigure}
    \hfill
    \begin{subfigure}[t]{0.19\linewidth}
        \centering
        \includegraphics[width=\linewidth]{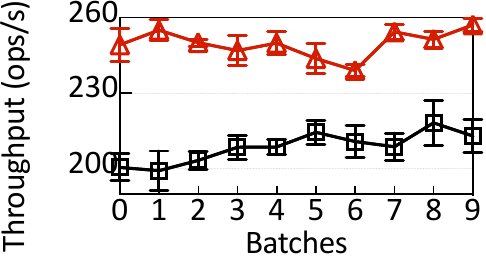}
        \vspace{-0.2in}
        \caption{CW}
    \end{subfigure}

    \vspace{-0.15in}
    \caption{{Throughput comparison with standard deviation over three runs under insertions.}}
    \label{fig:insert_ops}
     \vspace{-0.1in}
\end{figure*}

\begin{figure}[tbp]
    \centering
    \begin{subfigure}[t]{0.42\linewidth}
        \centering
        \includegraphics[width=\linewidth]{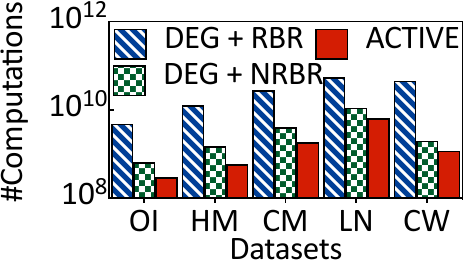}
        \vspace{-0.2in}
        \caption{Deletions}
    \end{subfigure}
    \hfill
    \begin{subfigure}[t]{0.42\linewidth}
        \centering
        \includegraphics[width=\linewidth]{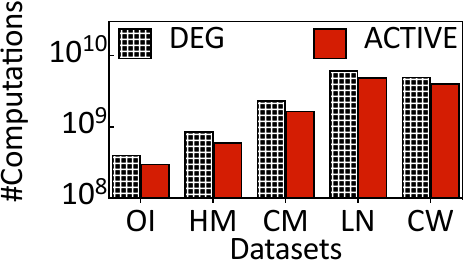}
        \vspace{-0.2in}
        \caption{Insertions}
    \end{subfigure}
    \vspace{-0.15in}
    \caption{{Distance computation comparison.}}
    \label{fig:cmp_count}
    \vspace{-0.25in}
\end{figure}

\stitle{{Query Latency under Varying $k$}}{.}
{
We measure the query latency for $k \in \{10,100,500,1000\}$ with Recall@$k$ fixed at $0.95$, as shown in Figure~\ref{fig:diffk}.
On LN, MR is omitted because Recall = 0.95 cannot be achieved within the reasonable latency budget.
Across all $k$ values, \oursys achieves substantially lower latency than MR in most cases, as MR misses many objects with small weighted distances during candidate generation. 
{MR can achieve recall comparable to that of \oursys only by substantially increasing its candidate budget (i.e., $ef_{\mathrm{search}}$), as shown in Figure~\ref{fig:diff_ef}.}
Meanwhile, \oursys maintains latency comparable to DEG while reducing AR computation time, as shown in Table~\ref{tab:construct}.
As $k$ increases, the latency advantage of \oursys over MR generally becomes more pronounced, while its relative latency performance against DEG also improves.
{For example, at $k=10$, the query latency ratios
of MR and DEG to \oursys range from $0.21$ to $2.35$
and from $0.63$ to $1.24$, respectively.
At $k=1000$, these ratios range from $2.13$ to $4.63$
and from $0.67$ to $1.29$, respectively.}
{
On HM, \oursys incurs higher search latency than DEG because its neighbors are more widely dispersed in Pareto-angle space, with the mean angular difference over all unordered neighbor pairs, averaged equally across nodes, reaching $16.07^\circ$ on HM, compared with $8.73^\circ$ on OI and $10.38^\circ$ on CM.
This requires a larger threshold to retain useful pruning comparisons. When $\theta_{th}$ exceeds $\pi/6$, \oursys achieves search performance competitive with DEG, as shown in Figure~\ref{fig:diff_theta_hm}.
A detailed analysis is provided in Appendix~\ref{sec:hm_threshold_sensitivity}.}
}

{Notably, on CW, MR outperforms \oursys and DEG when $k=10$.
This is because CW has low coordinate diversity, with only about $1993$ unique coordinates among $\sim$7.2M objects. In this case, all retrieved results share the same coordinate values. Consequently, the query results are dominated by single-vector retrieval over the text, effectively reducing the search to a single-vector search.
However, as $k$ increases, the retrieved results exhibit multiple distinct coordinate values, making the joint consideration of text and coordinate distances increasingly important.
}

\begin{figure*}[tbp]
    \centering
    \vspace{-0.5in}
    \begin{subfigure}[t]{0.35\linewidth}
        \centering
        \includegraphics[width=\linewidth]{expr/sec7/single_0601/delete_title.pdf}
    \vspace{-0.2in}
    \end{subfigure}

    \vspace{0.2em}

    \begin{subfigure}[t]{0.19\linewidth}
        \centering
        \includegraphics[width=\linewidth]{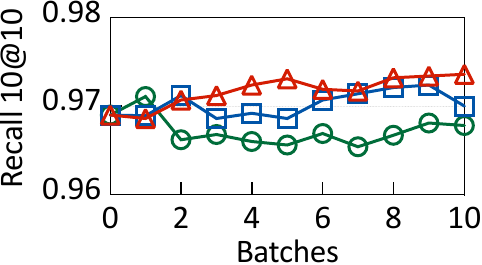}
    \vspace{-0.2in}
        \caption{OI}
    \end{subfigure}
    \hfill
    \begin{subfigure}[t]{0.19\linewidth}
        \centering
        \includegraphics[width=\linewidth]{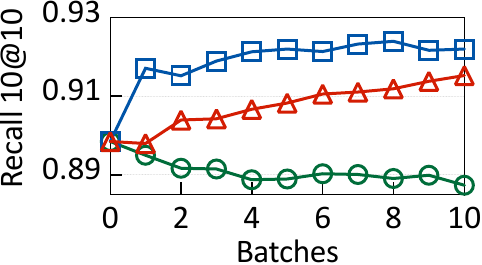}
    \vspace{-0.2in}
        \caption{HM}
    \end{subfigure}
    \hfill
    \begin{subfigure}[t]{0.19\linewidth}
        \centering
        \includegraphics[width=\linewidth]{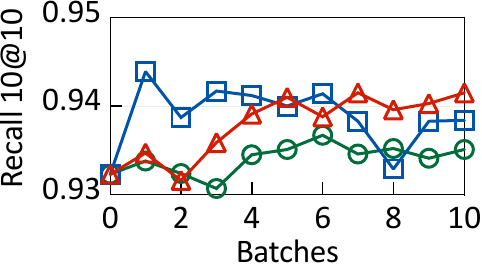}
    \vspace{-0.2in}
        \caption{CM}
    \end{subfigure}
    \hfill
    \begin{subfigure}[t]{0.19\linewidth}
        \centering
        \includegraphics[width=\linewidth]{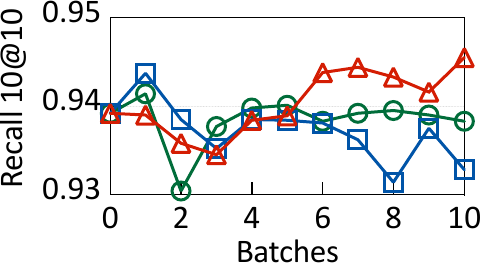}
    \vspace{-0.2in}
    \caption{LN}
    \end{subfigure}
    \hfill
    \begin{subfigure}[t]{0.19\linewidth}
        \centering
        \includegraphics[width=\linewidth]{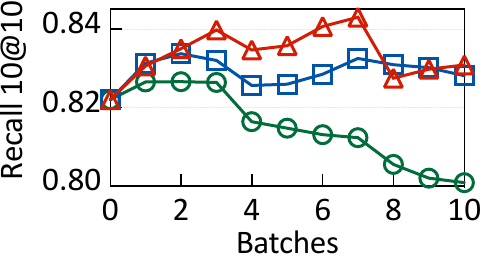}
    \vspace{-0.2in}
    \caption{CW}
    \end{subfigure}
    \vspace{-0.15in}
    \caption{{Search accuracy comparison under deletions.}}
    \label{fig:deletion_recall}
\end{figure*}

\begin{figure*}[tbp]
    \centering
    \vspace{-0.1in}

    \begin{subfigure}[t]{0.19\linewidth}
        \centering
        \includegraphics[width=\linewidth]{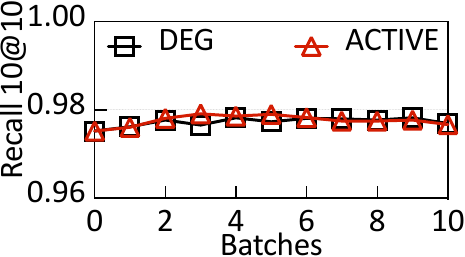}
        \vspace{-0.2in}
        \caption{OI}
    \end{subfigure}
    \hfill
    \begin{subfigure}[t]{0.19\linewidth}
        \centering
        \includegraphics[width=\linewidth]{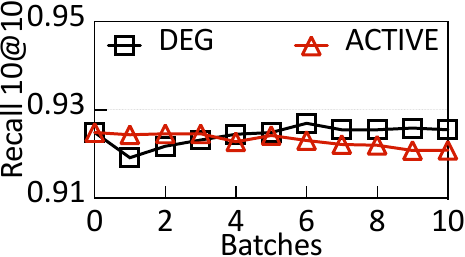}
        \vspace{-0.2in}
        \caption{HM}
    \end{subfigure}
    \hfill
    \begin{subfigure}[t]{0.19\linewidth}
        \centering
        \includegraphics[width=\linewidth]{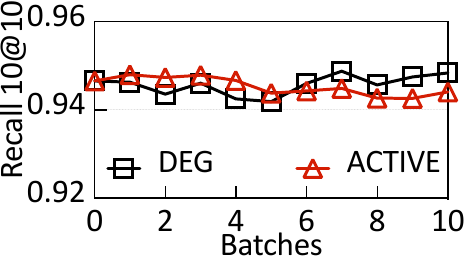}
        \vspace{-0.2in}
        \caption{CM}
    \end{subfigure}
    \hfill
    \begin{subfigure}[t]{0.19\linewidth}
        \centering
        \includegraphics[width=\linewidth]{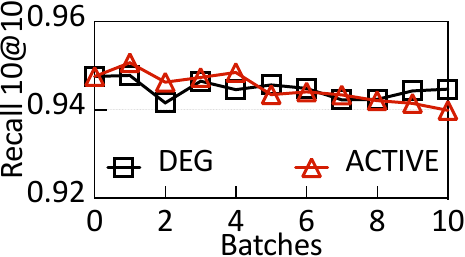}
        \vspace{-0.2in}
        \caption{LN}
    \end{subfigure}
    \hfill
    \begin{subfigure}[t]{0.19\linewidth}
        \centering
        \includegraphics[width=\linewidth]{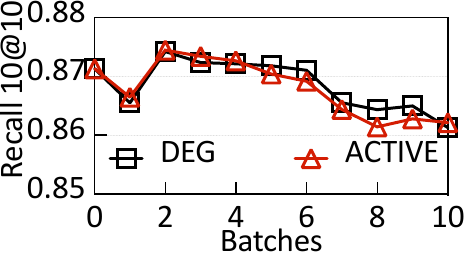}
        \vspace{-0.2in}
        \caption{CW}
    \end{subfigure}
    \vspace{-0.15in}
    \caption{{Search accuracy comparison under insertions.}}
    \label{fig:insert_recall}
     \vspace{-0.1in}
\end{figure*}

\begin{figure}[tbp]
    \centering

    \begin{subfigure}[t]{0.435\linewidth}
        \centering
        \includegraphics[width=\linewidth]{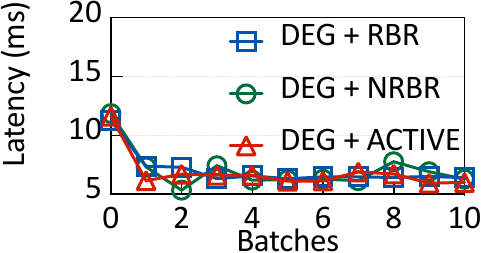}
        \vspace{-0.2in}
        \caption{Deletions}
    \end{subfigure}
    \hfill
    \begin{subfigure}[t]{0.435\linewidth}
        \centering
        \includegraphics[width=\linewidth]{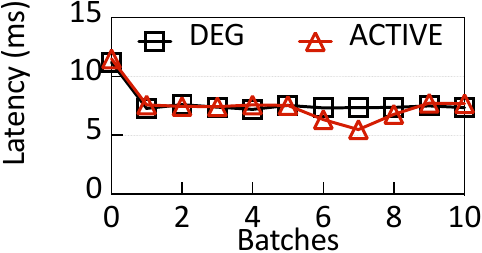}
        \vspace{-0.2in}
        \caption{Insertions}
    \end{subfigure}
    \vspace{-0.15in}
    \caption{Search latency comparison.}
    \label{fig:search_latency}
    \vspace{-0.1in}
\end{figure}

\stitle{{Impact of $\alpha$ Distributions}}{.}
{To reflect realistic query preferences that may occur in practice, we generate three Beta distributions that favor the second attribute, treat both attributes equally, and favor the first attribute. We evaluate the accuracy--efficiency trade-off on OI, as shown in Figure~\ref{fig:diff_alpha}. Pareto-frontier-based indices remain stable under all three distributions, whereas MR is highly sensitive to the $\alpha$ distribution and degrades sharply when $\alpha$ concentrates around $0.5$ (see Figure~\ref{fig:diff_alpha_0.5}).}

\vspace{-0.1in}
\subsection{Update Performance}
\label{subsec:expr_update}

We evaluate the update (i.e., deletion and insertion) performance of different methods on five real-world vector datasets, reporting the average results with standard deviation over three runs to mitigate system-level fluctuations.

\stitle{Deletion Throughput Comparison}.
{As shown in Figure~\ref{fig:delete_ops}, \oursys consistently achieves higher throughput than both RBR and NRBR across all datasets, improving deletion throughput by \delopsonerbrmin--$39.51\times$ over RBR and \delopsonenrbrmin--\delopsonenrbrmax$\times$ over NRBR.}
The key advantages of \oursys lie in its connectivity-loss-driven adaptive repair (\pfrepair) strategy and \acprune, both of which substantially reduce the computational overhead during deletions.
It is noteworthy that the throughput of \oursys is relatively low in the first batch, since the reverse graph needs to be explicitly initialized, whereas it can be maintained asynchronously in subsequent batches.

\stitle{Insertion Throughput Comparison}.
{As shown in Figure~\ref{fig:insert_ops}, \oursys improves insertion throughput by $1.20$--\insopsonemax$\times$ compared with the incremental construction of DEG.}
The improvement is attributed to \acprune, which reduces the number of distance computations by introducing Pareto-angle constraints.

\stitle{{Throughput Variability}}{.}
{As shown in Figure~\ref{fig:delete_ops} and Figure~\ref{fig:insert_ops},
the throughput gaps between \oursys and the baselines are consistently larger than the run-to-run variability indicated by the standard deviation error bars.}

\stitle{Distance Computation Comparison}.
We further compare the number of distance computations of \oursys against other methods across all datasets, as shown in Figure~\ref{fig:cmp_count}.
{Under 1\% deletions, \oursys consistently requires fewer computations than RBR and NRBR, reducing the computation count by \delcmponerbrmin$\times$--$38.70\times$ over RBR and by $1.70\times$--\delcmponenrbrmax$\times$ over NRBR.}
This is attributed to the fact that \pfrepair adaptively selects different repair strategies for each affected node instead of applying a fixed repair strategy, even when the connectivity loss ${CL}$ of the affected node is small.
{Under 1\% insertions, \oursys reduces the number of computations compared to DEG, achieving a reduction of $1.23\times$--\inscmponemax$\times$, thanks to \acprune.}

\subsection{Index Quality}
\label{subsec:expr_search}

Although all designs in \oursys substantially improve update performance, it is necessary to evaluate whether \oursys can maintain graph index quality during updates. We therefore evaluate search performance under 1\% deletions and 1\% insertions.

\stitle{Search Accuracy Comparison under deletions}.
As shown in Figure~\ref{fig:deletion_recall},
compared with RBR, \oursys achieves comparable recall {(with a Recall@10 gap ranging from $-1.23\%$ to $+0.49\%$)} across most datasets, indicating that our local adaptive repair strategy can closely match the effectiveness of global reconstruction for affected nodes. 
{On HM, RBR achieves higher search accuracy than \oursys, but with $30.28\times$ slower deletion throughput. The recall gap arises because local repair in \oursys cannot always recover Pareto frontiers as accurate as those of RBR.}
{Compared with NRBR, \oursys improves search accuracy by \delrecallonenrbrmin--$2.03\%$.}

\stitle{Search Accuracy Comparison under Insertions}.
Figure~\ref{fig:insert_recall} shows that \oursys achieves recall comparable to that of DEG across all datasets {(with a Recall@10 gap ranging from $-0.13\%$ to $+0.04\%$)}, indicating that our proposed \acprune improves the AR computation efficiency while maintaining search accuracy closely matching the baseline.

\stitle{Search Latency Comparison}.
To further assess the impact of different update methods on search performance, we measure the average search latency on the largest dataset, LN, separately under deletions and insertions.
As shown in Figure~\ref{fig:search_latency}, \oursys achieves search latency comparable to that of other methods and remains consistently stable under both deletions and insertions.

These results show that \acprune and \pfrepair improve update performance while keeping index quality comparable to the baselines.

\subsection{{Update Stability}}
{We also evaluate the stability of update performance under streaming data on OI. Specifically, we set the sliding window size to cover the first $50\%$ of the dataset and use these nodes to build the initial index. In each batch of updates, the window slides forward by $1\%$ of the dataset, i.e., we insert $1\%$ of the remaining data and delete $1\%$ of the data from the beginning of the window.}

\begin{figure}[tbp]
    \centering
    \begin{subfigure}[t]{0.9\linewidth}
        \centering
        \includegraphics[width=\linewidth]{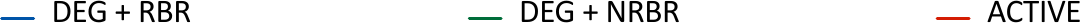}
        \vspace{-0.2in}
    \end{subfigure}
    \begin{subfigure}[t]{0.42\linewidth}
        \centering
        \includegraphics[width=\linewidth]{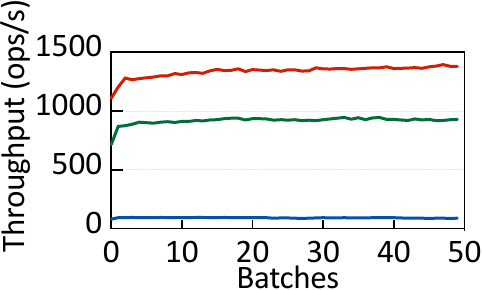}
        \vspace{-0.2in}
        \caption{Update}
        \label{fig:us_update}
    \end{subfigure}
    \hfill
    \begin{subfigure}[t]{0.42\linewidth}
        \centering
        \includegraphics[width=\linewidth]{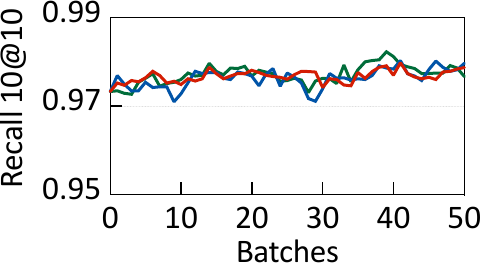}
        \vspace{-0.2in}
        \caption{Search}
        \label{fig:us_search}
    \end{subfigure}
    \vspace{-0.15in}
    \caption{{Update stability comparison.}}
    \label{fig:update_stability}
    \vspace{-0.1in}
\end{figure}

\stitle{{Long-Term Update Stability}}{.}
{We repeat the sliding-window update process $50$ times, by which point the entire index has been replaced and a completely new index has been constructed. Figure~\ref{fig:us_update} shows that all methods maintain stable update performance, and \oursys consistently retains a clear advantage. We further evaluate search accuracy after each update, as shown in Figure~\ref{fig:us_search}. \oursys achieves search accuracy comparable to other baselines.}

\begin{figure}[tbp]
    \centering
    \begin{subfigure}[t]{0.9\linewidth}
        \centering
        \includegraphics[width=\linewidth]{revision/pdf/title.pdf}
    \vspace{-0.2in}
    \end{subfigure}

    \vspace{0.2em}
    \begin{subfigure}[t]{0.33\linewidth}
        \centering
        \includegraphics[width=\linewidth]{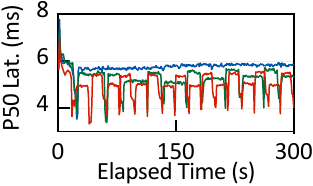}
        \vspace{-0.2in}
        \caption{P50 latency}
    \end{subfigure}
    \hspace{-0.02\linewidth}
    \begin{subfigure}[t]{0.33\linewidth}
        \centering
        \includegraphics[width=\linewidth]{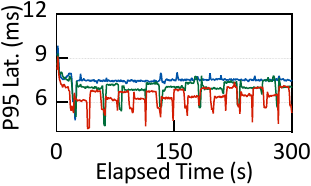}
        \vspace{-0.2in}
        \caption{P95 latency}
    \end{subfigure}
    \hspace{-0.02\linewidth}
    \begin{subfigure}[t]{0.33\linewidth}
        \centering
        \includegraphics[width=\linewidth]{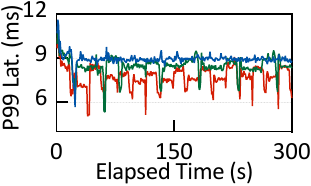}
        \vspace{-0.2in}
        \caption{P99 latency}
    \end{subfigure}
    \vspace{-0.15in}
    \caption{{Query latency under a mixed workload.}}
    \label{fig:p99_oi}
    \vspace{-0.15in}
\end{figure}

\stitle{{Mixed Reader/Writer Workload}}{.}
{
We evaluate queries under concurrent updates on the OI index, with $16$ update threads continuously performing sliding-window updates and $16$ query threads issuing search requests in parallel.
We record the p50, p95, and p99 query latencies under continuously batched updates and show the first 300 seconds of the trace for clarity. As shown in Figure~\ref{fig:p99_oi}, \oursys achieves lower query latency than the other baselines under continuous batch updates. %
Notably, all methods exhibit periodic latency drops. This occurs because, between consecutive update batches and between insertion and deletion within a single batch, read--write lock contention is briefly released and query threads face less competition for CPU and memory bandwidth.}

\subsection{Scalability}
\label{subsec:scal}
We evaluate deletion and insertion performance on the OI dataset under different batch sizes to assess scalability with respect to update granularity.

\begin{figure}[tbp]
    \centering

    \begin{subfigure}[t]{0.4\linewidth}
        \centering
        \includegraphics[width=\linewidth]{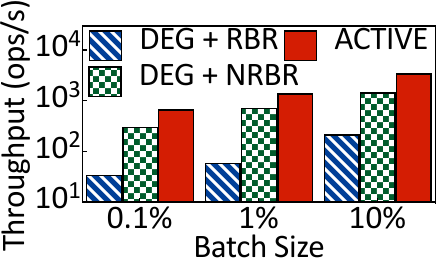}
        \vspace{-0.2in}
        \caption{Deletion throughput}
        \label{fig:diff_del_ops}
    \end{subfigure}
    \hfill
    \begin{subfigure}[t]{0.4\linewidth}
        \centering
        \includegraphics[width=\linewidth]{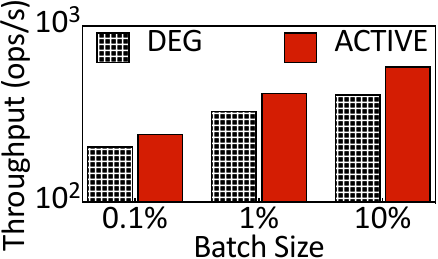}
        \vspace{-0.2in}
        \caption{Insertion throughput}
        \label{fig:diff_ins_ops}
    \end{subfigure}

    \ \

    \begin{subfigure}[t]{0.4\linewidth}
        \centering
        \includegraphics[width=\linewidth]{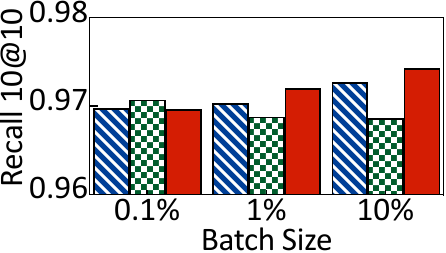}
        \vspace{-0.2in}
        \caption{Deletion accuracy}
        \label{fig:diff_del_ac}
    \end{subfigure}
    \hfill
    \begin{subfigure}[t]{0.4\linewidth}
        \centering
        \includegraphics[width=\linewidth]{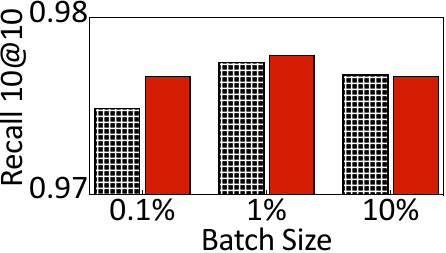}
        \vspace{-0.2in}
        \caption{Insertion accuracy}
        \label{fig:diff_ins_ac}
    \end{subfigure}

    \vspace{-0.15in}
    \caption{Varying batch size of deletions and insertions.}
    \label{fig:varing_size}
    \vspace{-0.15in}
\end{figure}

\stitle{Throughput under Different Batch Sizes}.
As shown in Figure~\ref{fig:diff_del_ops} and \ref{fig:diff_ins_ops}, when the batch size increases from 0.1\% to 10\% of the dataset, all methods achieve higher update throughput. This is because larger batch updates amortize fixed system overheads, thereby improving overall throughput.
Notably, \oursys consistently outperforms all baselines across different batch sizes under both deletions and insertions, indicating that our proposed \acprune and \pfrepair provide a consistent performance advantage across a wide range of update granularity.

\stitle{Search Accuracy under Different Batch Sizes}.
Figure~\ref{fig:diff_del_ac} shows that as batch size increases, recall of \oursys and RBR improves while NRBR degrades, indicating that NRBR fails to repair connectivity loss under large-batch deletions. \oursys achieves recall comparable to RBR across all batch sizes.
Figure~\ref{fig:diff_ins_ac} shows that \oursys consistently achieves search accuracy comparable to DEG across all batch sizes, validating the effectiveness of \acprune.

\subsection{Parameter Sensitivity}
\label{subsec:expr_theta}
In \oursys, update performance is mainly affected by three parameters, namely the Pareto angle threshold $\theta_{th}$ in \acprune, and the connectivity loss thresholds ${CL}_1$ and ${CL}_2$ in \pfrepair.
We next study the impact of these parameters on update throughput and recall by varying them, and motivate the choice of the default settings.

\begin{figure}[tbp]
    \centering
    \vspace{-0.05in}
    \begin{subfigure}[t]{0.7\linewidth}
        \centering
        \includegraphics[width=\linewidth]{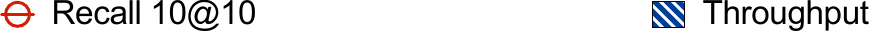}
        \vspace{-0.2in}
    \end{subfigure}
    \begin{subfigure}[t]{0.47\linewidth}
        \centering
        \includegraphics[width=\linewidth]{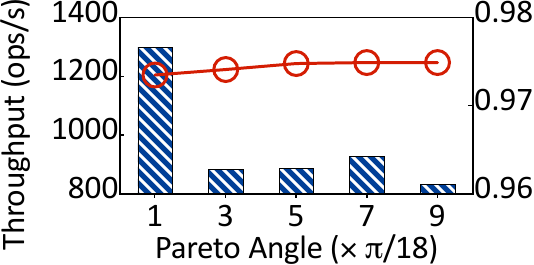}
        \vspace{-0.2in}
        \caption{Deletions}
    \end{subfigure}
    \hfill
    \begin{subfigure}[t]{0.45\linewidth}
        \centering
        \includegraphics[width=\linewidth]{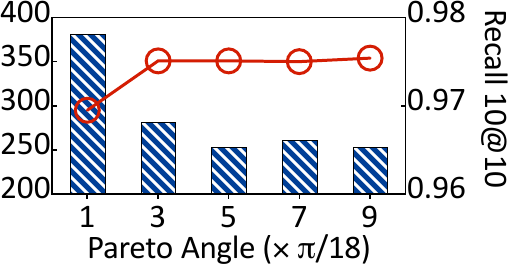}
        \vspace{-0.2in}
        \caption{Insertions}
    \end{subfigure}
    \vspace{-0.15in}
    \caption{{Sensitivity to the Pareto angle threshold $\theta_{th}$.}}
    \label{fig:angle_expr}
     \vspace{-0.15in}
\end{figure}

\stitle{Pareto Angle Threshold}.
In \acprune, a larger $\theta_{th}$ allows each candidate node to be compared with more previously retained nodes, resulting in lower efficiency of the AR computation.
To evaluate its impact, we conduct a sensitivity analysis on OI under 1\% deletions and insertions by varying $\theta_{th}$ among $\pi/18$, $\pi/6$, $5\pi/18$, $7\pi/18$, and $\pi/2$, as shown in Figure~\ref{fig:angle_expr}.
{The results show that $\theta_{th}=\pi/18$ achieves the highest throughput on both insertion and deletion, with only a small loss in Recall@10 relative to larger $\theta_{th}$ values.}
Therefore, we set $\theta_{th}$ to $\pi/18$ by default.

\begin{figure}[tbp]
    \centering
    \vspace{-0.55in}
    \begin{subfigure}[t]{0.47\linewidth}
        \centering
        \includegraphics[width=\linewidth]{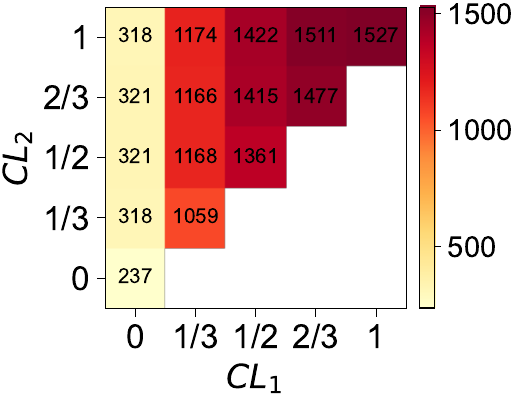}
        \vspace{-0.25in}
        \caption{Deletion throughput}
    \end{subfigure}
    \hfill
    \hspace{-0.1in}
    \begin{subfigure}[t]{0.47\linewidth}
        \centering
        \includegraphics[width=\linewidth]{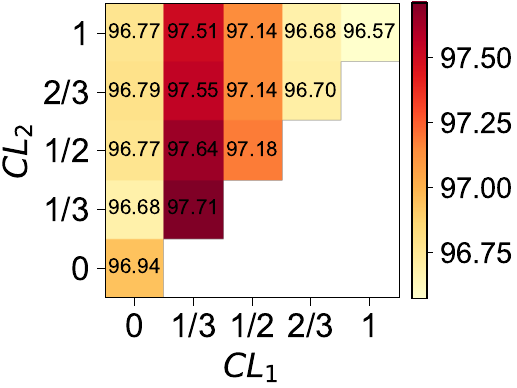}
        \vspace{-0.25in}
        \caption{Search accuracy (\%)}
    \end{subfigure}
    \vspace{-0.15in}
    \caption{{Sensitivity to connectivity loss thresholds.}}
    \label{fig:delta_expr}
    \vspace{-0.05in}
\end{figure}

\stitle{Connectivity Loss Thresholds}.
{We further evaluate joint $\mathrm{CL}_1\times\mathrm{CL}_2$ settings on OI.
Figure~\ref{fig:delta_expr} shows that the default setting $\mathrm{CL}_1=1/3$ and $\mathrm{CL}_2=1/2$ provides a consistent throughput--accuracy trade-off.
Compared with the highest-recall setting ($\mathrm{CL}_1=\mathrm{CL}_2=1/3$), it improves deletion throughput by $10.28\%$ with only $0.07\%$ lower Recall@10. Compared with the highest-throughput setting ($\mathrm{CL}_1=\mathrm{CL}_2=1$), it improves Recall@10 by $1.07\%$. These results support $\mathrm{CL}_1=1/3$ and $\mathrm{CL}_2=1/2$ as robust defaults.}

\stitle{{Tuning Guidance}}{.}
{We recommend using the default parameters of \oursys as the first choice.
They are empirical default values obtained from experiments across different datasets, different update batch sizes, different $\alpha$ distributions, and different update patterns. For datasets or workloads that require fine-tuning, one can run experiments similar to those in Figure~\ref{fig:angle_expr} and Figure~\ref{fig:delta_expr} on randomly sampled dual-vector data to select more suitable parameters.
Specifically, increasing $\theta_{th}$ improves search accuracy but increases the cost of active range computation. For $\mathrm{CL}_1$ and $\mathrm{CL}_2$, we recommend adjusting $\mathrm{CL}_1$ first to obtain a better throughput--accuracy trade-off for the current workload.}

\subsection{Performance Gain}
\label{subsec:expr_perf_gain}

\begin{figure}[t]
    \centering
    \begin{subfigure}[t]{0.9\linewidth}
        \centering
        \includegraphics[width=\linewidth]{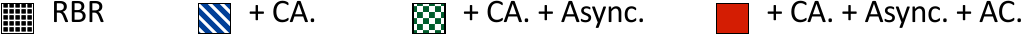}
        \vspace{-0.2in}
    \end{subfigure}
    \begin{subfigure}[t]{0.32\linewidth}
        \centering
        \includegraphics[width=\linewidth]{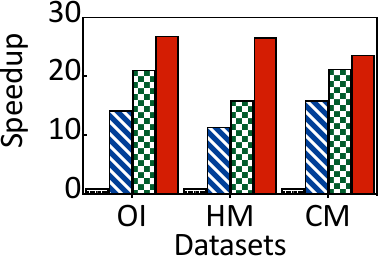}
        \vspace{-0.2in}
        \caption{0.1\% Deletion}
    \end{subfigure}
    \hfill
    \begin{subfigure}[t]{0.32\linewidth}
        \centering
        \includegraphics[width=\linewidth]{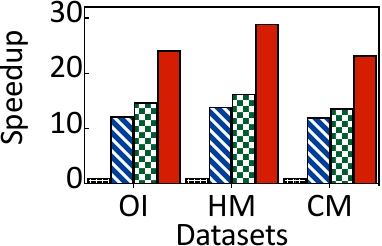}
        \vspace{-0.2in}
        \caption{1\% Deletion}
    \end{subfigure}
    \hfill
    \begin{subfigure}[t]{0.32\linewidth}
    \centering
    \includegraphics[width=\linewidth]{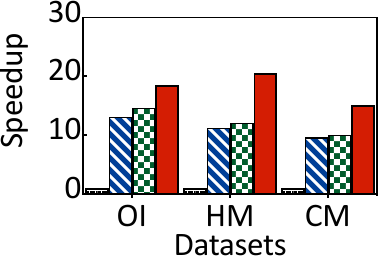}
        \vspace{-0.2in}
    \caption{10\% Deletion}
    \end{subfigure}
    \vspace{-0.15in}
    \caption{Performance gain.}
    \label{fig:perform_gain}
     \vspace{-0.05in}
\end{figure}

To evaluate the contribution of each core component to deletion performance, we start from RBR and progressively add the proposed components on three datasets under 0.1\%, 1\%, and 10\% deletion batch sizes (see Figure~\ref{fig:perform_gain}), reporting normalized throughput relative to RBR. Each component yields consistent speedups across batch sizes. Specifically, we first introduce the connectivity-loss-driven adaptive repair algorithm (+C.A.), which adaptively repairs affected nodes according to the degree of connectivity loss, achieving average speedups of 13.85$\times$, 12.77$\times$, and 11.34$\times$ under the three batch sizes, respectively. We then further incorporate an asynchronously maintained reverse graph (+Async.), increasing the average speedups to 19.46$\times$, 14.95$\times$, and 12.27$\times$, respectively, by avoiding scanning the entire graph topology. Finally, we introduce the angle-constrained AR computation algorithm (+A.C.), improving the average speedups to 25.76$\times$, 25.50$\times$, and 18.05$\times$, respectively.

Notably, the asynchronously maintained reverse graph provides significant performance gains under the 0.1\% batch size. Its main benefit is eliminating full graph scans, whereas the cost of in-memory access is relatively small.

\subsection{{Ablation Study}}
\label{subsec:expr_ablation}

{To justify the three connectivity-loss levels and the separate treatment of the first Pareto-frontier layer, we conduct experiments to validate these design choices.}

\begin{figure}[tbp]
    \centering
    \vspace{-0.05in}
    \begin{subfigure}[t]{0.51\linewidth}
        \centering
        \includegraphics[width=\linewidth]{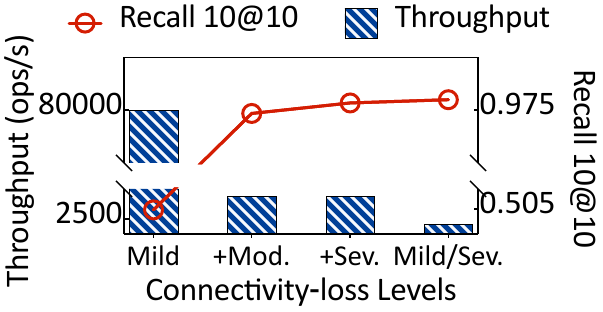}
        \vspace{-0.15in}
        \caption{{Ablation study of \pfrepair}}
        \label{fig:ablation_pfrepair}
    \end{subfigure}
    \hspace{0.02\linewidth}
    \begin{subfigure}[t]{0.43\linewidth}
        \centering
        \includegraphics[width=\linewidth]{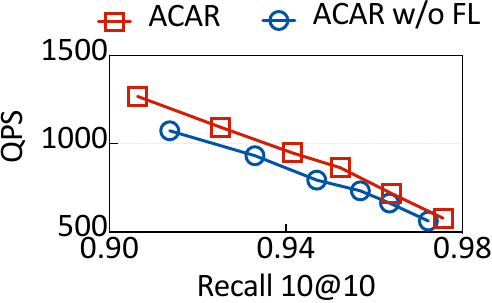}
        \vspace{-0.15in}
        \caption{Ablation study of \acprune}
        \label{fig:ablation_acprune}
    \end{subfigure}

    \vspace{-0.15in}
    \caption{{Ablation studies of \pfrepair and \acprune.}}
    \label{fig:ablation_for_alg}
    \vspace{-0.15in}
\end{figure}

\stitle{{Ablation of \pfrepair}}{.}
{
{We compare deletion throughput and Recall@10 for four threshold settings on OI after 90 batches of 1\% deletions.}
The first uses $\mathrm{CL}_1=\mathrm{CL}_2=1$ (only Mild), where all affected nodes are repaired as mild loss. The second uses $\mathrm{CL}_1=1/3$ and $\mathrm{CL}_2=1$ (+Moderate), which yields only mild and moderate loss. The third uses $\mathrm{CL}_1=1/3$ and $\mathrm{CL}_2=1/2$ (+Severe), which is the full three-level \pfrepair.
As shown in Figure~\ref{fig:ablation_pfrepair}, the full strategy (+Severe) achieves the best overall trade-off.
{Although the first strategy achieves higher throughput, it fails to maintain graph connectivity under sustained deletions.}
{Compared with the forth strategy, which uses $\mathrm{CL}_1=1/3$ and $\mathrm{CL}_2=1/3$ (Mild/Severe), \pfrepair achieves 7.51\% higher deletion throughput with only a 0.03\% decrease in recall.}
}

\stitle{{Ablation of \acprune}}{.}
{To examine whether the first Pareto-frontier layer requires special treatment (FL), we compare the full \acprune with a variant without this treatment (\acprune w/o FL), which applies the same angle-constrained procedure uniformly to all layers, including the first layer. As shown in Figure~\ref{fig:ablation_acprune}, the full \acprune consistently achieves a better accuracy-efficiency trade-off.}

\vspace{-0.05in}
\subsection{{Profiling Breakdown of Update Time}}
\label{sec:profiling}

\begin{figure}[tbp]
    \centering
    \vspace{-0.55in}
    \begin{subfigure}[t]{0.9\linewidth}
        \centering
        \includegraphics[width=\linewidth]{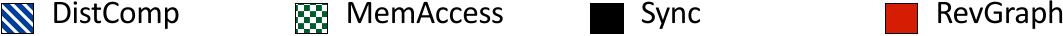}
        \vspace{-0.2in}
    \end{subfigure}
    \vspace{-0.1in}
    \begin{subfigure}[t]{0.42\linewidth}
        \centering
        \includegraphics[width=\linewidth]{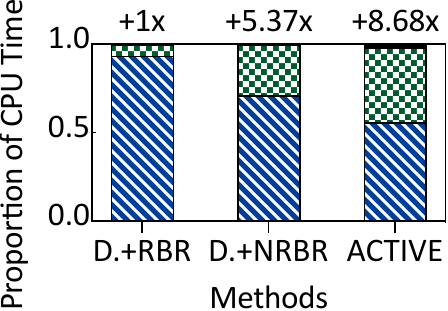}
        \vspace{-0.2in}
        \caption{OI}
    \end{subfigure}
    \hfill
    \begin{subfigure}[t]{0.4\linewidth}
        \centering
        \includegraphics[width=\linewidth]{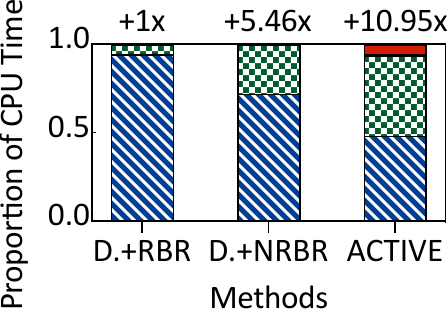}
        \vspace{-0.2in}
        \caption{HM}
    \end{subfigure}
    \caption{{Profiling breakdown of update time.}}
    \label{fig:profiling}
     \vspace{-0.05in}
\end{figure}

{To understand the practical sources of speedup beyond asymptotic complexity, we profile the update procedure on the OI and HM datasets under $1\%$ batch updates (deletions and insertions), with the time breakdown shown in Figure~\ref{fig:profiling}.
The results indicate that \oursys achieves its speedup primarily by reducing distance computation, while the additional reverse-graph maintenance overhead remains small.
The remaining cost components are comparable to those of the baselines, which is consistent with our analysis.}

\section{Related Work}
\label{sec:relatedwork}

\stitle{Hybrid Search}.
To meet the demands of increasingly complex real-world applications, various hybrid search paradigms have been proposed, including filtered ANNS~\cite{filter-anns,rw-sv4,vbase,navix} and multi-vector ANNS~\cite{milvus}. 
Filtered ANNS associates each object with both vector representations and structured attributes, supporting approximate nearest neighbor search under attribute constraints such as numerical ranges~\cite{range_f1,range_f2,range_f3,range_f4,unify,DIGRA,wow}, temporal conditions~\cite{temp_f}, and categorical filters~\cite{Filtered-DiskANN,label-f1,vec-flow}. 
Multi-vector ANNS represents each object with multiple embeddings and jointly optimizes similarity across vector spaces~\cite{onesparse,deg,IGP,plaid,MUVERA,ColBERT}. 
These paradigms extend traditional ANNS by incorporating structured or multi-view semantic information.

\stitle{Dual-Vector Query}.
Existing dual-vector query approaches fall into two categories: preference-specific multi-index methods~\cite{milvus,multi-recall} and preference-agnostic unified index methods~\cite{deg}. 
The former fails to maintain high search accuracy under arbitrary query preferences, while the latter achieves a better accuracy–efficiency trade-off but does not support efficient %
updates.

\stitle{{Chamfer-based Multi-Vector Retrieval (CMVR)}}{.}
{The similarity between objects in CMVR~\cite{MUVERA} is measured by $\operatorname{CHAMFER}(Q,P)=\sum_{q_i \in Q}\max_{p_j \in P}\langle q_i,p_j\rangle$, where $P$ and $Q$ denote the document and query vector sets, respectively.
CMVR can capture finer-grained semantic information within a single feature than traditional $L_2$ or cosine distance measures applied to single-vector representations. %
This setting differs substantially from DVQ in both the retrieval task and the similarity measure, and therefore CMVR methods, such as MUVERA~\cite{MUVERA}, cannot be used directly in DVQ tasks.}

\stitle{{Graph-based Index Updates}}{.}
{In single-vector graph-based indices, dynamic updates~\cite{ip-diskann,wolverine,greator,odinann} have been extensively studied. Insertions are typically performed via incremental construction~\cite{FreshDiskANN,greator,wolverine}, and recent work such as ParlayANN~\cite{parlayann} further accelerates this process through scalable and deterministic parallel batch insertion. Deletions are handled by either global reinsertion~\cite{HNSW,IPGM} or local neighbor reconnection~\cite{FreshDiskANN,greator}.}

\vspace{-0.05in}
\section{Conclusion}
\label{sec:concl}

We propose \oursys, a dynamic Pareto-frontier-based unified DVQ index that supports efficient updates while maintaining comparable retrieval quality under arbitrary query preferences. To accelerate index construction and insertions, we design an angle-constrained active range computation algorithm, \acprune, which reduces distance computations via Pareto-angle-based constraints. To support efficient updates, we further propose a connectivity-loss-driven adaptive repair algorithm (\pfrepair) that selects different repair strategies based on the connectivity loss of affected nodes.
Experiments demonstrate that \oursys achieves a strong trade-off between retrieval quality and update throughput across datasets.

\clearpage
\bibliographystyle{ACM-Reference-Format}
\bibliography{sample}

\clearpage
\appendix

\section{Applicability to Other Dual Attributes}

\stitle{Datasets with Additional Dual Attributes}.
\oursys is not limited to image--text data, but supports arbitrary combinations of data types.
To further examine the applicability of \oursys, we add two geo-tagged datasets, as shown in Table~\ref{tab:rs_datasets}. The first is Regions (RS)~\cite{regions}, a multilingual geotagged social-media corpus collected from Twitter, where each text post is embedded using BERT~\cite{devlin-etal-2019-bert}. The second is MP16-Pro (MP)~\cite{g3}, a geo-tagged image collection from Flickr, where each image is embedded into a 768-dimensional vector using DINOv2 ViT-B/14~\cite{oquab2024dinov2}. We conduct both static search and dynamic update experiments on RS and MP.

\begin{table}[h]
\centering
\vspace{-0.1in}
\caption{Dataset statistics. $D_x$ and $D_y$ denote the vector dimensions of the $x$- and $y$-attributes, respectively. Each dataset contains 1,000 query vectors.}
\label{tab:rs_datasets}
\vspace{-0.1in}

\begin{tabular}{lccrl}
\toprule
Dataset & $D_x$ & $D_y$ & \#Vectors & Contents \\
\midrule
Regions (RS)  & 768 & 2 & 614,000   & Text-Coord.  \\
MP16-Pro (MP) & 768 & 2 & 1,000,000 & Image-Coord. \\
\bottomrule
\end{tabular}
\vspace{-0.15in}
\end{table}

\begin{figure}[H]
    \centering
    \vspace{-0.1in}
    \includegraphics[width=0.85\linewidth]
        {expr/sec7/single/static_title.pdf}

    \vspace{0.2em}

    \begin{subfigure}[t]{0.42\linewidth}
        \centering
        \includegraphics[width=\linewidth]
            {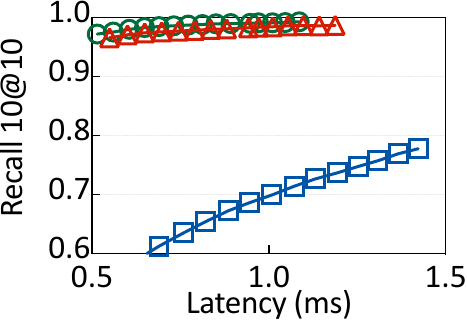}
        \vspace{-0.2in}
        \caption{Regions}
        \label{fig:static_rs}
    \end{subfigure}
    \hfill
    \begin{subfigure}[t]{0.42\linewidth}
        \centering
        \includegraphics[width=\linewidth]
            {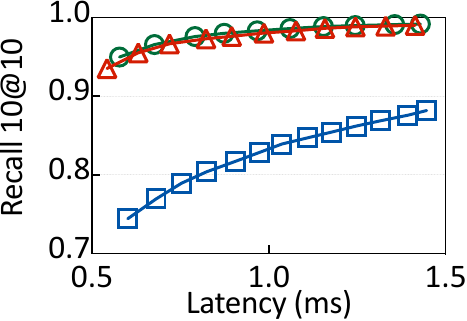}
        \vspace{-0.2in}
        \caption{MP16-Pro}
        \label{fig:static_mp}
    \end{subfigure}

    \vspace{-0.15in}
    \caption{Comparison of the trade-offs between search accuracy and latency. The top left is better.}
    \label{fig:static_search_geo}
    \vspace{-0.15in}
\end{figure}

\stitle{Static Search Performance}.
The accuracy-efficiency trade-off results of \oursys, DEG, and MR on RS and MP are shown in Figure~\ref{fig:static_search_geo}. Under the same search latency, the Recall@10 of \oursys is on average $0.83\%$ and $0.32\%$ lower than that of DEG on RS and MP, respectively. Meanwhile, \oursys substantially outperforms MR, improving Recall@10 by $33.10\%$ on RS and by $15.17\%$ on MP under the same latency. These results show that Pareto-frontier-based indices remain effective for text-coordinate and image-coordinate attribute pairs, and that \oursys preserves search quality close to DEG while clearly surpassing multi-index retrieval.

\begin{figure}[H]
    \centering
    \vspace{-0.1in}
    \begin{subfigure}[t]{0.42\linewidth}
        \centering
        \includegraphics[width=\linewidth]
            {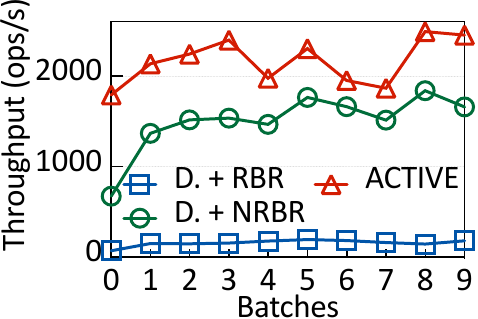}
        \vspace{-0.2in}
        \caption{Regions}
        \label{fig:delete_ops_rs}
    \end{subfigure}
    \hfill
    \begin{subfigure}[t]{0.42\linewidth}
        \centering
        \includegraphics[width=\linewidth]
            {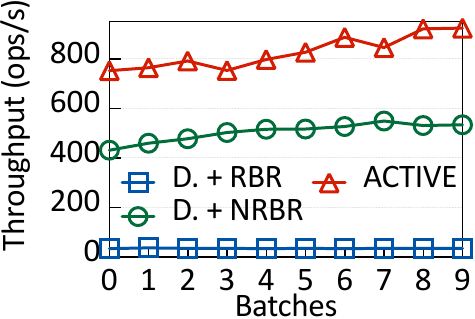}
        \vspace{-0.2in}
        \caption{MP16-Pro}
        \label{fig:delete_ops_mp}
    \end{subfigure}

    \vspace{-0.1in}
    \caption{Deletion throughput comparison.}
    \label{fig:delete_ops_geo}
    \vspace{-0.15in}
\end{figure}

\begin{figure}[H]
    \centering
    \begin{subfigure}[t]{0.42\linewidth}
        \centering
        \includegraphics[width=\linewidth]
            {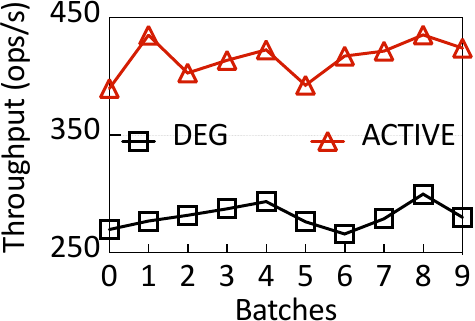}
        \vspace{-0.2in}
        \caption{Regions}
        \label{fig:insert_ops_rs}
    \end{subfigure}
    \hfill
    \begin{subfigure}[t]{0.42\linewidth}
        \centering
        \includegraphics[width=\linewidth]
            {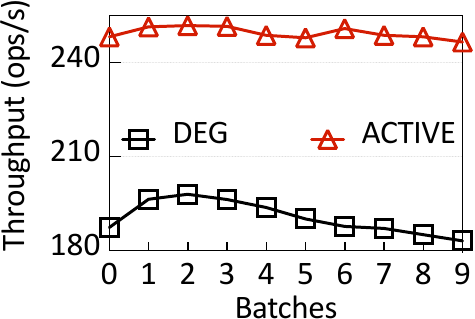}
        \vspace{-0.2in}
        \caption{MP16-Pro}
        \label{fig:insert_ops_mp}
    \end{subfigure}

    \vspace{-0.1in}
    \caption{Insertion throughput comparison.}
    \label{fig:insert_ops_geo}
    \vspace{-0.15in}
\end{figure}

\stitle{Update Throughput}.
We evaluate the update throughput of \oursys on RS and MP under 1\% batch deletions and insertions, as shown in Figures~\ref{fig:delete_ops_geo} and~\ref{fig:insert_ops_geo}. For deletions, \oursys consistently achieves higher throughput than both RBR and NRBR, improving the average deletion throughput by $13.78$--$23.88\times$ over RBR and $1.42$--$1.64\times$ over NRBR. For insertions, \oursys improves the average throughput by $1.31$--$1.48\times$ over DEG.

\begin{figure}[H]
    \centering
    \vspace{-0.15in}
    \begin{subfigure}[t]{0.42\linewidth}
        \centering
        \includegraphics[width=\linewidth]
            {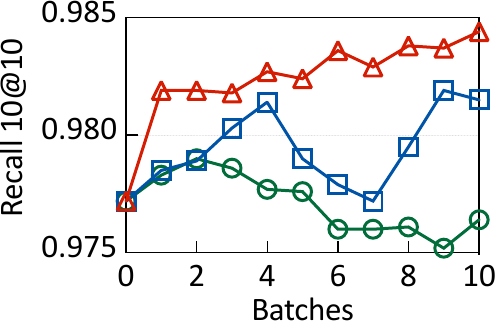}
        \vspace{-0.2in}
        \caption{Regions}
        \label{fig:delete_recall_rs}
    \end{subfigure}
    \hfill
    \begin{subfigure}[t]{0.42\linewidth}
        \centering
        \includegraphics[width=\linewidth]
            {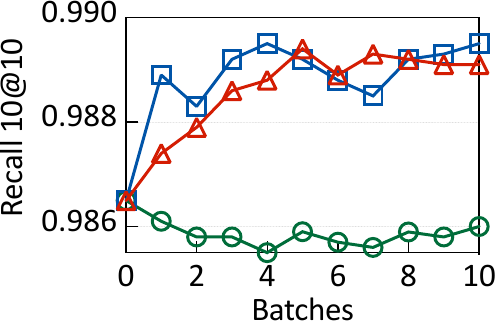}
        \vspace{-0.2in}
        \caption{MP16-Pro}
        \label{fig:delete_recall_mp}
    \end{subfigure}

    \vspace{-0.1in}
    \caption{Recall comparison under deletions.}
    \label{fig:delete_recall_geo}
    \vspace{-0.15in}
\end{figure}

\begin{figure}[H]
    \centering
    \vspace{-0.15in}
    \begin{subfigure}[t]{0.42\linewidth}
        \centering
        \includegraphics[width=\linewidth]
            {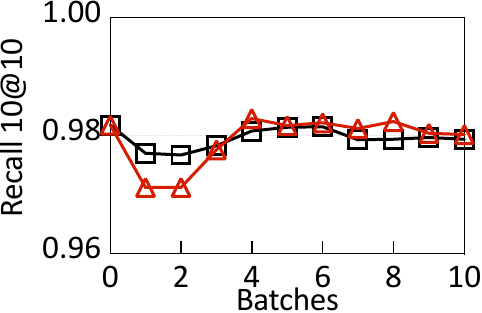}
        \vspace{-0.2in}
        \caption{Regions}
        \label{fig:insert_recall_rs}
    \end{subfigure}
    \hfill
    \begin{subfigure}[t]{0.42\linewidth}
        \centering
        \includegraphics[width=\linewidth]
            {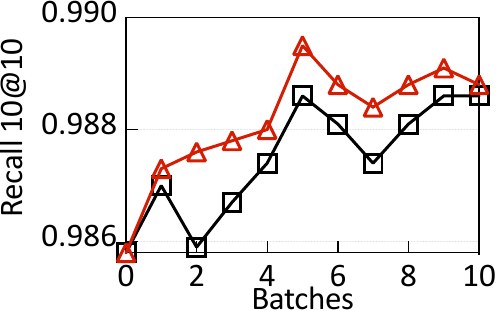}
        \vspace{-0.2in}
        \caption{MP16-Pro}
        \label{fig:insert_recall_mp}
    \end{subfigure}

    \vspace{-0.1in}
    \caption{Recall comparison under insertions.}
    \label{fig:insert_recall_geo}
    \vspace{-0.1in}
\end{figure}

\stitle{Search Accuracy under Updates}.
We further evaluate the search accuracy of \oursys under 1\% batch deletions and insertions on RS and MP, as shown in Figures~\ref{fig:delete_recall_geo} and~\ref{fig:insert_recall_geo}. For deletions, \oursys achieves Recall@10 comparable to that of RBR and $0.30\%$--$0.58\%$ higher than that of NRBR on average. For insertions, \oursys maintains Recall@10 comparable to that of DEG.

\vspace{-0.05in}
\section{Analysis of Affected Nodes}
\label{subsec:ana_aff_nodes}

\begin{figure}[H]
    \centering
    \begin{subfigure}[t]{0.7\linewidth}
        \centering
        \includegraphics[width=\linewidth]{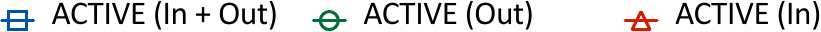}
        \vspace{-0.2in}
    \end{subfigure}
    \vspace{-0.1in}
    \begin{subfigure}[t]{0.4\linewidth}
        \centering
        \includegraphics[width=\linewidth]{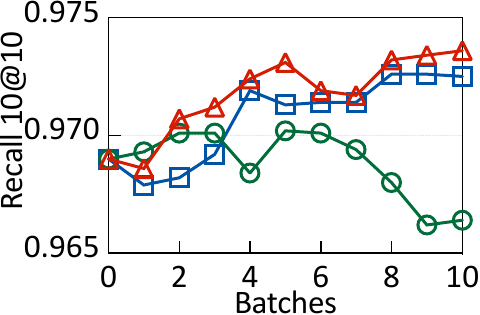}
        \vspace{-0.2in}
        \caption{Search accuracy}
    \end{subfigure}
    \hfill
    \begin{subfigure}[t]{0.4\linewidth}
        \centering
        \includegraphics[width=\linewidth]{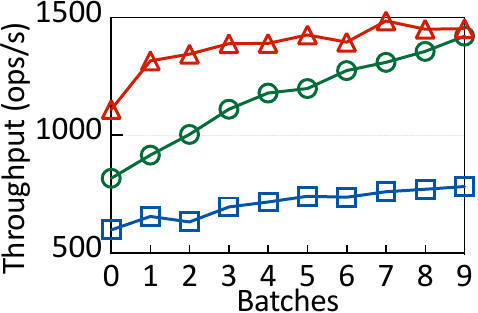}
        \vspace{-0.2in}
        \caption{Deletion Throughput}
    \end{subfigure}
    \caption{Impact of different affected node definitions.}
    \label{fig:diff_repair}
     \vspace{-0.15in}
\end{figure}

By default, \oursys treats in-neighbors of deleted nodes as affected nodes, since deletions mainly cause connectivity loss of in-neighbors.
We argue that repairing only in-neighbors is sufficient to largely restore connectivity.
To validate this, we compare \oursys (In) with two variants on OI under 1\% batch deletions. ACTIVE (Out) repairs only out-neighbors, while ACTIVE (In+Out) repairs both in-neighbors and out-neighbors of deleted nodes.
As shown in Figure~\ref{fig:diff_repair}, ACTIVE (In) consistently achieves higher update throughput and better search accuracy than other repair strategies.

\section{Sensitivity Analysis of the Pareto Angle Threshold $\theta_{{th}}$}
\label{sec:hm_threshold_sensitivity}

To understand why HM is more sensitive to the Pareto angle threshold
$\theta_{{th}}$, we analyze the Pareto angle dispersion of neighbors across all objects in the dataset.

\begin{figure}[!htbp]
    \centering
    \begin{subfigure}[t]{0.32\linewidth}
        \centering
        \includegraphics[width=\linewidth]
            {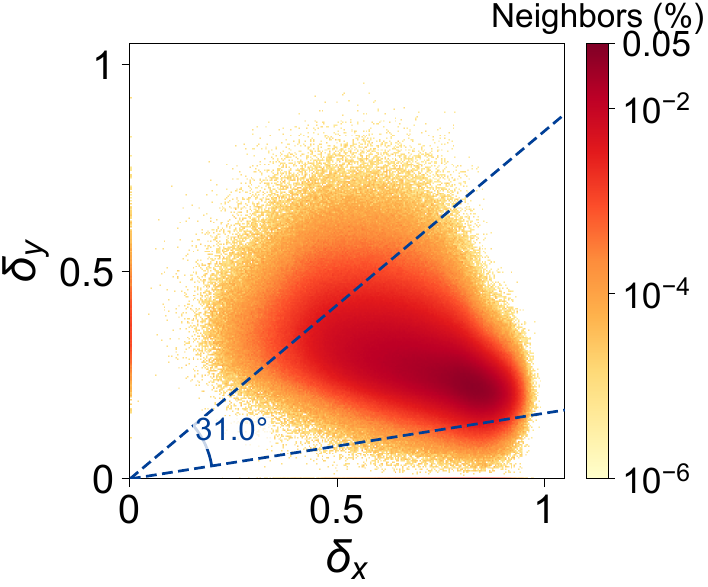}
        \caption{OpenImage}
        \label{fig:OI_phi}
    \end{subfigure}
    \hfill
    \begin{subfigure}[t]{0.32\linewidth}
        \centering
        \includegraphics[width=\linewidth]
            {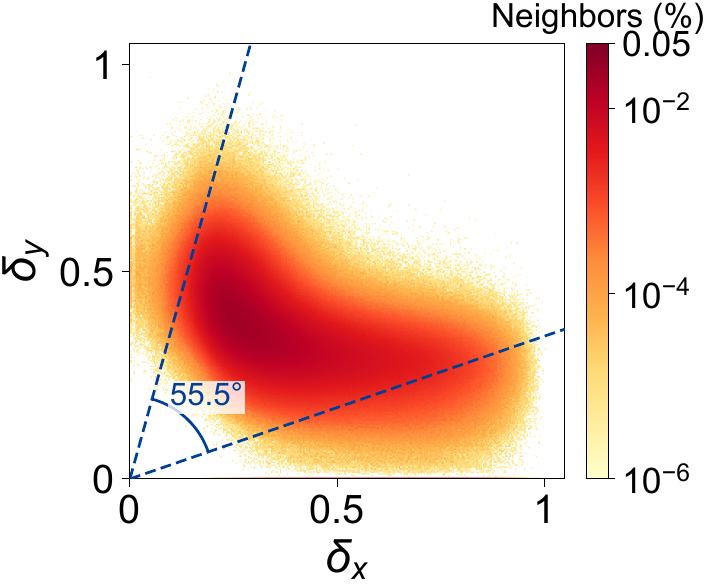}
        \caption{Howto100M}
        \label{fig:hm_phi}
    \end{subfigure}
    \hfill
    \begin{subfigure}[t]{0.32\linewidth}
        \centering
        \includegraphics[width=\linewidth]
            {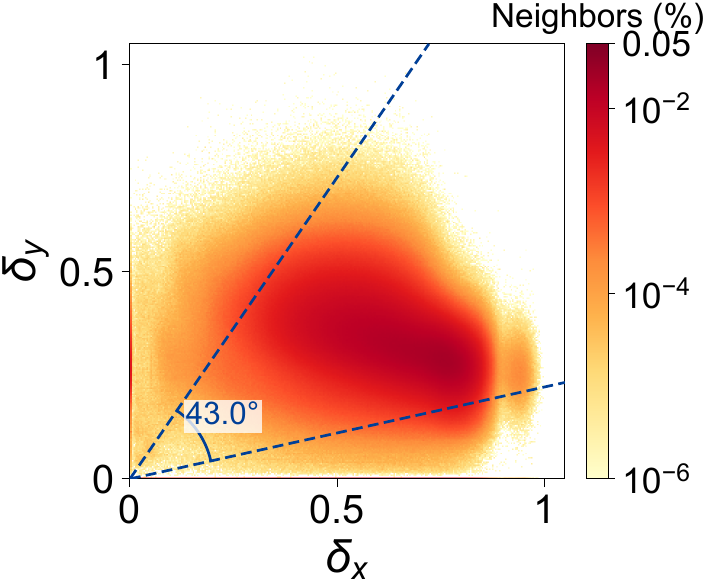}
        \caption{CC3M}
        \label{fig:cm_phi}
    \end{subfigure}
    \caption{Distribution of dual-attribute distances between nodes
    and their neighbors. The blue dashed rays delimit the shortest
    continuous Pareto-angle interval containing 95\% of the neighbor
    pairs. The angular width of each interval is annotated.}
    \label{fig:density}
\end{figure}

\stitle{Pareto Angle Dispersion}.
We analyze the distribution of dual-attribute distances between each node $u$ and its neighbors, as shown in Figure~\ref{fig:density}.
Specifically, for each node $u$, we collect the distance pairs $(\delta_x(u,v), \delta_y(u,v))$ for all its neighbors $v$ and aggregate them across all nodes to characterize the overall neighbor distribution of each dataset.
HM exhibits a wider Pareto-angle interval containing 95\% of the neighbor pairs, spanning $55.5^\circ$, compared with $31.0^\circ$ for OI and $43.0^\circ$ for CM.
To further quantify neighbor dispersion, we compute the mean Pareto angle difference over all unordered neighbor pairs of each node and then average these values equally across nodes.
Under the same maximum neighbor count $R$, the resulting mean is $16.07^\circ$ on HM, compared with $8.73^\circ$ on OI and $10.38^\circ$ on CM.
These results indicate that the neighbors of individual nodes in HM tend to be more widely spread and dispersed in the Pareto angular space.

{The dispersion is related to the consistency of object distributions across the two attribute spaces. On OI and CM, two objects that are similar in one attribute tend to be similar in the other, whereas this relationship is much weaker on HM.}
Consequently, a small threshold may exclude more potentially useful pruning comparisons in HM, thereby affecting neighbor selection quality and search performance.

\clearpage

\clearpage

\end{document}